\documentclass[a4paper,12pt]{article}

\pdfoutput=1

\usepackage[paper=letterpaper,margin=1in]{geometry}

\usepackage{amsmath,amssymb,amsthm,amsfonts,epsfig,cite,setspace,bigstrut,longtable,array,url,color}

\usepackage{hyperref}

\newcommand{\todo}[1]{{\bf ?????!!!! #1 ?????!!!!}\marginpar{$\Longleftarrow$}}

\newcommand{\nn}{\nonumber}
\newcommand{\tr}{\mathop{\rm Tr}}
\newcommand{\comment}[1]{}

\newcommand{\cL}{{\cal L}}
\newcommand{\cM}{{\cal M}}
\newcommand{\cW}{{\cal W}}
\newcommand{\cN}{{\cal N}}
\newcommand{\cH}{{\cal H}}

\newcommand{\cO}{{\cal O}}

\newcommand{\cT}{{\cal T}}

\newcommand{\IN}{\mathbb{N}}
\newcommand{\IP}{\mathbb{P}}
\newcommand{\IQ}{\mathbb{Q}}

\newcommand{\IR}{\mathbb{R}}
\newcommand{\IC}{\mathbb{C}}

\newcommand{\II}{\mathbb{I}}
\newcommand{\IZ}{\mathbb{Z}}
\newcommand{\re}{{\rm~Re}}
\newcommand{\im}{{\rm~Im}}

\newcommand{\rad}{{\rm Rad}}

\renewenvironment{thebibliography}[1]{%
\begin{oldthebibliography}{#1}%
\setlength{\parskip}{0ex}%
\setlength{\itemsep}{0ex}%
}%
{%
\end{oldthebibliography}%
}

\newtheorem{theorem}{\bf THEOREM}
\newtheorem{proposition}{\bf PROPOSITION}
\newtheorem{lemma}{\bf LEMMA}

\newcommand{\setall}{\setcounter{equation}{0}
  \setcounter{theorem}{0}}

\begin{document}
\begin{titlepage}

  ~\\
  \vspace{4mm}

\begin{center}
{\Large \bf Yang-Mills Theory and the ABC Conjecture}
\medskip

\vspace{4mm}

{\large Yang-Hui He}$^{1,2,3}$,
{\large Zhi Hu}$^{1,4}$,
{\large Malte Probst}$^{1,5}$,
{\large James Read}$^{2}$

\vspace{1mm}

\renewcommand{\arraystretch}{0.5} 
{\small
{\it
\begin{tabular}{rl}
${}^{1}$ &
Department of Mathematics, City University, London, EC1V 0HB, UK\\
${}^{2}$ &
Merton College, University of Oxford, OX14JD, UK\\
${}^{3}$ &
School of Physics, NanKai University, Tianjin, 300071, P.R.~China\\
${}^{4}$ &
School of Mathematics, University of Science and Technology of China,\\
&Wu Wen Tsun Key Laboratory of Mathematics, Chinese Academy of Science\\
&Hefei, Anhui, 230026, P.~R.~China \\
${}^{5}$ &
Department of Physics, Universit\"at Heidelberg, 69117, Germany\\
\end{tabular}
}
~\\
~\\
~\\
\begin{tabular}{cc}
hey@maths.ox.ac.uk, &  halfask@mail.ustc.edu.cn, \\
m.probst@stud.uni-heidelberg.de, & james.read@merton.ox.ac.uk
\end{tabular}
}
\renewcommand{\arraystretch}{1.5} 

\end{center}

\vspace{10mm}

\begin{abstract}
  We establish a precise correspondence between the ABC Conjecture and $\cN=4$ super-Yang-Mills theory.
  This is achieved by combining three ingredients: (i) Elkies' method of mapping ABC-triples to elliptic curves in his demonstration that ABC implies Mordell/Faltings; (ii) an explicit pair of elliptic curve and associated Belyi map given by Khadjavi-Scharaschkin; and (iii) the fact that the bipartite brane-tiling/dimer model for a gauge theory with toric moduli space is a particular {\it dessin d'enfant} in the sense of Grothendieck.
  We explore this correspondence for the highest quality ABC-triples as well as large samples of random triples. The Conjecture itself is mapped to a statement about the fundamental domain of the toroidal compactification of the string realization of $\cN=4$ SYM.
\end{abstract}

\end{titlepage}

\tableofcontents

\vspace{30mm}

\section{Introduction}\setall
The ABC Conjecture \cite{OM}, having resided at the heart of number theory for some decades, remains as tantalizing today as ever. Deceptively simple to state -- much like many of the deepest results in mathematics -- its importance lies in its plethora of implications, ranging from L-functions of elliptic curves, to the distribution of primes, to the asymptotic form of Fermat's Last Theorem (cf.~a dedicated page at \cite{implications}).
In the last decade, Mochizuki \cite{mochi} has both intrigued the professional community and captivated the public imagination in his announcement of a proof using some extraordinary and yet to be understood constructs; an attempted decipherment thereof constituted a productive 2016 Oxford workshop \cite{conrad}.

In a parallel and seemingly unrelated vein, one of the chief protagonists of modern theoretical physics is $\cN=4$ Yang-Mills theory (SYM), a finite, maximally supersymmetric, conformal quantum field theory in $3+1$-dimensions. 
Since its emergence some 4 decades ago \cite{Brink:1976bc}, its central position in a diversity of profound physics, ranging from S-duality, to AdS/CFT, to exact scattering amplitudes, has been unquestioned.
Over the last 10 years, a programme had been established to understand gauge theories whose vacuum moduli space is a toric variety -- the prototype of which being $\cN=4$ SYM -- via dimer models/brane tilings on a torus \cite{Feng:2000mi,Franco:2005rj}.

That the above two contemporaneous directions of research should have any intersection seems fantastical, were it not for the fact that elliptic curves should act as the point d'appui in both cases.
In the former, this is unsurprising, since at least when Frey mapped Fermat to an elliptic curve which now bears his name; in the latter, the torus on which the tiling realizes the embedding of $\cN=4$ SYM into string theory has its complex structure rigidly fixed when the dimer manifests as a dessin d'enfant, or upon $a$-maximization.

Using the technique of Elkies \cite{elkies} in showing that ABC implied the Mordell Conjecture/Faltings' Theorem, Khadjavi and Scharaschkin \cite{KS,KS1} mapped any ABC-triple onto a rational point on any one of three types of elliptic curves, including, in particular, the curve of $j$-invariant 0, viz., $y^2 = x^3 + D^2$ with a Belyi map $\beta = (y+D)/(2D)$ for a cube-free integer $D$.
This pair of curve and Belyi map is precisely the one which describes $\cN=4$ SYM, through a trivalent dessin which renders a hexagonal tiling of the torus \cite{Jejjala:2010vb,Hanany:2011ra}.

This pleasantly unexpected identification, that the ABC Conjecture and $\cN=4$ SYM should be encoded into {\it exactly} the same curve with the same Belyi map, thus naturally comprises the commencement of our investigations, and we shall delve into how statements regarding ABC triples can be transformed into the physics.
Indeed, this approach of recasting problems in number theory into physical systems dates back to Hilbert and Polya a century ago in rephrasing the Riemann Hypothesis in terms of Hermitian observables -- which has since blossomed into a dynamic field itself -- and has without a doubt yielded a fruitful dialogue. It seems here too we have a wonderful situation in which there is a natural interplay between physics and number theory.

The outline of the paper is as follows.
We begin in \S2 with a review of the ABC Conjecture in the context of rational points on elliptic curves, distinguishing a particular family with $j$-invariant 0 and an associated Belyi map.
Next, in \S3, we show how this same curve and map encode $\cN=4$ SYM, both in embedding in string theory and purely as a field theory.
We then study in \S4 how high quality ABC-triples are distributed on the brane tiling, and indeed, how the triples are distributed in general, before moving to conclusions and prospects in \S\ref{s:conc}.

\section{ABC Conjecture, Elliptic Curves and Belyi Maps}\setall
Let us begin by recalling the statement of the ABC Conjecture.
There are many versions thereof and we will present perhaps the most standard \cite{OM}.
The reader is also referred to a nice pedagogical account in \cite{MM}.
Let $(a,b,c)$ be coprime integers such that $a+b=c$ which, for convenience, we will take to be all positive so that $c = \max(a,b,c)$.
Next, recall the definition of the radical $\rad(x)$ of a (positive) integer $x$ with prime factorization $x = \prod\limits_i p_i^{a_i}$: this is the product of the prime factors taken only once, i.e., $\rad(x) = \prod\limits_i p_i$.
The ABC Conjecture then asserts that $\rad(abc)$ cannot be too small compared to $c$ in the following sense.
\begin{quote}
  {\rm \bf ABC Conjecture}: We have the following four equivalent formulations for positive coprime triples of integers with $a+b=c$
  \begin{enumerate}
  \item For every $\epsilon > 0$, there exists only finitely many such $(a,b,c)$ such that $c > \rad(abc)^{1+\epsilon}$;
  \item For every $\epsilon > 0$, there exists a constant $K_\epsilon$ such that for all such triples $(a,b,c)$, we have $c < K_\epsilon \rad(abc)^{1+\epsilon}$;
  \item Define the {\em quality} to be $q(a,b,c) := \frac{\log(c)}{\log(\rad(abc))}$, then there exist only finitely many such $(a,b,c)$ with $q(a,b,c) > 1 + \epsilon$ for all $\epsilon>0$;
  \item $\limsup\limits_{c \to \infty} q(a,b,c) = 1$.
  \end{enumerate}
\end{quote}

That $\epsilon$ is indispensable is easily seen as follows.
Consider $(a,b,c) = (1, 2^{6n}-1, 2^{6n})$.
Then $\rad(abc)=2\rad(b) \le 2\cdot3\rad(b/9)$ since $b$ is divisible by $64-1$, and hence 9.
In turn this is less than or equal to $3 \cdot b/9$, so that  $\rad(abc) \le 2b/3 = 2/3(c-1) < c$ and we have an infinite family of triples with $c > \rad(abc)$, i.e., infinitely many triples with $q > 1$.
However, with the addition of $\epsilon$, the conjecture is valid.

In \S2.5.4 of the classic book \cite{LZ} on graphs on Riemann surfaces, especially on Grothendieck's {\it dessins d'enfants}, the authors discuss how a particular Belyi function from $\IP^1  \to \IP^1$, viz., $f(x) = \frac{64x^3}{(x+9)^3(x+1)}$, upon the substitution of $x = r/s$ for coprime integers $(r,s)$ produces $(a,b,c)$ of unusually high quality.
Indeed, the ABC Conjecture, amongst its many consequences, is inextricably linked to the theory of elliptic curves \cite{mochi}; in particular, it was shown \cite{elkies} that it implied the celebrated Mordell Conjecture.

\subsection{Elliptic Curves and Belyi Maps}
That the ABC Conjecture is intimately related to elliptic curves is well known.
In fact, an equivalent statement to ABC is the so-called {\bf Szpiro Conjecture}, which phrases the problem entirely in terms of an elliptic curve $E$ over $\IQ$ and compares the minimal discriminant $\Delta$ and the conductor $f$: that for any $\epsilon > 0$, there exists a constant $C_\epsilon$ such that $|\Delta| \le C_\epsilon f^{6 + \epsilon}$.

In \cite{elkies}, Elkies generalized the notion of the radical and defined an appropriate norm so that the ABC Conjecture can be stated for any number field $K$.
In particular, he showed that the conjecture is powerful enough to imply the Mordell Conjecture (Falting's Theorem) that the number of rational points over any Riemann surface of genus greater than one is finite \cite{MF}.
Crucial to his argument is the construction of a meromorphic function ramified on the Riemann surface at only three points, a so-called Belyi function, whose rudiments we now briefly recall;
the reader is referred to the standard books \cite{LZ,Leila} and a rapid introduction and brief review in \cite{He:2012js}.

\subsubsection{Belyi's Theorem and Dessins d'Enfants}
Let $\Sigma$ be an algebraic curve (Riemann surface) over $\IC$ of genus $g$.
Then we have the remarkable statement of G.~Belyi dating to as late as 1979 (the proof involves techniques known to the 19th century) that
\begin{theorem} [Belyi]
$\Sigma$ has a model over the algebraic closure $\overline{\mathbb{Q}}$ of the rationals {\bf if and only if} there exists a holomorphic covering $\beta : \Sigma \longrightarrow \mathbb{P}^1(\mathbb{C})$ as a rational map, ramified over only three points, which may be taken as $\left\{0,1,\infty\right\}$ by a M\"{o}bius transformation.
\end{theorem}
We refer to the combination $\left(\Sigma,\beta\right)$ as a {\it Belyi pair}.
Explicitly, $\Sigma$ can be written in standard hyper-elliptic form in affine coordinates $(x,y)$, with $f(x)$ a degree $2g+1$ or $2g+2$ polynomial, and in the rational map, any expression in $y^2$ will be reduced to a polynomial in $x$ via the definition of the hyper-elliptic curve (the denominator can also be multiplied by a conjugate to remove $y$ entirely):
\begin{equation}
(\Sigma_{x,y}, \beta(x,y)) = \left(y^2 = f(x) \ , \  \frac{P\left(x\right)+R\left(x\right)y}{Q\left(x\right)} \right) \ ,
\end{equation}
with $P(x)$, $Q(x)$ and $R(x)$ some polynomials.

Subsequently, Grothendieck \cite{Leila} realized this as a {\it bipartite graph} embedded into $\Sigma$ by setting all pre-images of $0$ as, say, black nodes and those of $1$ as, say, white nodes; such a graph he called a {\bf dessin d'enfant}, or a child's drawing.
Next, the pre-image of any continuous curve from 0 to 1 will become segments which connect the black to the white nodes only (hence bipartite). Due to the Riemann-Roch Theorem, the pre-images of $\infty$ do not constitute an independent degree of freedom and are in fact 1-1 to interiors of polygons formed by the segments.
The valency of each node is given by the {\it ramification index} of $\beta$ in local coordinates around the pre-image.
The calculation of the ramification indices follows the methodology of, say, \cite{Hanany:2011ra}. It is helpful to introduce the {\it total derivative}, which is the derivative to be used when considering the order of vanishing (i.e., ramification) at the branch points when restricted to $\Sigma$. Defining $F\left(x,y\right) = y^2 - f\left(x\right)$, which must vanish identically on the curve, we have
\begin{equation}
\frac{d}{dx} = \frac{\partial}{\partial x} - \frac{\partial_x F}{\partial_y F}\frac{\partial}{\partial y} \ ,
\end{equation}
which is valid at the points where $x$ is a good local coordinate when $\partial_y F \neq 0$. Alternatively, we can use
\begin{equation}
\frac{d}{dy} = \frac{\partial}{\partial y} - \frac{\partial_y F}{\partial_x F}\frac{\partial}{\partial x},
\end{equation}
which is valid when $\partial_x F \neq 0$ and thus $y$ is a good local coordinate.
Finally, near the point $\left(\infty,\infty\right)$, where a good coordinate is $\epsilon$ with $x = 1/\epsilon^2$ and $y = 1/\epsilon^d$, where $d$ is the degree of the polynomial $f\left(x\right)$, the total derivative can be written as:
\begin{equation}
\frac{d}{d\epsilon} = -2y \frac{\partial}{\partial x} - dx^2\frac{\partial}{\partial y} \ ,
\end{equation}
so that if $\beta\left(\infty\right) = \infty$, then this derivative is understood to be acting on $1/\beta$.

With these in hand, we need only to follow a straightforward routine. If $\left(x_0^i,y_0^i\right)$ is a preimage of 0, then its ramification $r_0\left(i\right)$ is defined to be such that
$\left.\frac{d^k}{dx^k}\right|_{\left(x_0^i,y_0^i\right)} \beta\left(x,y\right) = 0$
for all $k = 0,1,2,\dots,r_0\left(i\right)-1$, where $k=0$ is just evaluation.
That is, it is the order of vanishing of $\beta$ along $\Sigma$.
We then follow a similar procedure to calculate $r_1\left(i\right)$ and $r_\infty(i)$.
Thus, to draw the dessin, to the $m$-th preimage of 0, we associate a black node with valency $r_0(m)$ (i.e., $r_0(m)$ edges), and to the $n$th preimage of 1, we associate a white node with valency $r_1(n)$.
We connect only black nodes to white nodes and vice versa, thereby forming a face, which is in fact a $(2r_\infty(k))-$gon.
We can record the ramification structure, known as the {\bf passport}, as
$\begin{Bmatrix}
r_0(1),r_0(2),\dots,r_0(B) \\
r_1(1),r_1(2),\dots,r_1(W) \\
r_\infty(1),r_\infty(2),\dots,r_\infty(I) 
\end{Bmatrix}$.
Finally, Riemann-Roch implies the following constraints on the passport:
\begin{equation}\label{RR}
d = \sum\limits_{i=1}^B r_0(i) = \sum\limits_{i=1}^W r_1(i) = \sum\limits_{i=1}^I r_\infty(i) \ , \qquad
2g - 2 = d - (B+W+I),
\end{equation}

\subsubsection{Belyi, Mordell \& ABC }
Now, for any ABC-triple, it suffices to know the ratio
\begin{equation}
r := c / b \in \IQ
\end{equation}
considered as a point in $\IP^1(\IQ)$, the projective line over $\IQ$.
Restoring, for the moment, the possibility of ($a,b,c)$ being negative integers as well, then since none of $(a,b,c)$ is zero, $r$ can be any point on $\IP^1(\IQ)$ except $\{0,1,\infty\}$.
In this way any $(a,b,c)$ triple is encoded into a single point on $\IP^1$ over $\IQ$.
Thus, ABC-triples can be mapped to an algebraic curve and an associated Belyi map in the ensuing.
As mentioned earlier, this method is due to Elkies \cite{elkies}, and we will shortly study some results of Khadjavi and Scharaschkin \cite{KS,KS1} which present explicit elliptic curves producing high quality ABC-triples.
Henceforth, we will continue to adhere to our convention that $(a,b,c)$ are positive integers so that our ratio $r > 1$ and we will reach a subset of relevant rational points on both the elliptic curve and the image $\IP^1(\IQ)$.

Consider an algebraic curve $\Sigma$ with a Belyi map $\beta : \Sigma \rightarrow \IP^1$.
Any $P \in \Sigma$ outside of $\beta^{-1}(\{0,1,\infty\})$ gives an ABC-triple, namely
\begin{equation}
  \beta(P) = r = c/b \ .
\end{equation}
We will write $q(P)$ as the quality $q(a,b,c)$ of the triple thus created.

Clearly, to obtain an infinite sequence of ABC-triples from $\Sigma$, we are forced to have the genus of $\Sigma$ being only 0 or 1 by Mordell/Faltings; in particular, we will focus on $g=1$ \cite{KS}.
Since one can state the ABC Conjecture as $\limsup\limits_{c \to \infty} q(a,b,c) = 1$, it behooves us to attempt to find infinite sequences of triples of high quality.
In a sense, the ABC Conjecture is a refinement of Mordell in that whilst on $g=1$ there might be an infinite number of rational points, there should be only a finite number with quality $q > 1+ \epsilon$ for every $\epsilon >0$.

For example, the highest quality so far is produced by Eric Reyssat (cf.~de Smit's ABC page \cite{abcpage}), viz.,
\begin{equation}\label{ER}
  2 + 3^{10} \cdot 109 = 23^5 \ , \quad
  q = \log(23^5) / \log(2 \cdot 3 \cdot 23 \cdot 109) \simeq 1.62991 \ .
\end{equation}
This corresponds to the Belyi pair
\begin{equation}
  \left(
  \Sigma = \{y^2 = x^3 + D x \} \ , \beta = -x^2/D
  \right) \ ,
  \quad D = -90252 = - 2^2 \cdot 3^2 \cdot 23 \cdot 109 \ .
\end{equation}
The affine point $(x,y) = (S/d^2, T/d^3)$ with $d = 3^2$, $S = 2 \cdot 23^3$ and $T = 2^2\cdot 23^2$, is mapped to $\beta = -x^2/D = \frac{23^5}{3^{10} \cdot 109} = c/b$, whence the triple $(a,b,c) = \left( 2 , \  3^{10} \cdot 109 , \  23^5 \right)$.

\subsection{High Quality and Belyi Maps}
Using the method of \cite{elkies}, Khadjavi and Scharaschkin \cite{KS1,KS} produce the following useful result:
\begin{theorem} (Khadjavi-Scharaschkin) {\bf Any} ABC-triple can be constructed from {\bf all of the following} situations of Belyi pairs $(E,\beta)$ of elliptic curve $E$ and Belyi map $\beta : E \to \IP^1$:
  \begin{enumerate}
  \item $(E,\beta) = (y^2 = x^3 + D x, -\frac{x^2}{D} = \frac{c}{b})$ for $D$ a fourth-power-free integer.
  \item $(E,\beta) = (y^2 = x^3 + D^2, \frac{y+D}{2D} = \frac{c}{b})$ for $D$ a cube-free integer.
  \item $(E,\beta) = (y^2 = x^3 + D, \frac{y^2}{D} = \frac{c}{b})$ for $D$ a sixth-power-free integer.
  \end{enumerate}
  \label{thm:KS}
\end{theorem}
In the following section we will see how the above theorem yields connections between ABC-triples and gauge theory.

That any ABC-triple can be reduced to a point on each of the curves is straight-forward and is worth expounding here.
For Type 1, let $c - b = a$ be the ABC-triple and multiply through by $a^3c^2$. Setting $x = ac$, $y = a^2 c$ and $D = -a^2bc$ gives the curve $y^2 = x^3 + D x$ with $-x^2/D = a^2c^2/(-a^2bc) = c/b$ as required. Now, if $D$ had a fourth-power factor, say $k^4 \in \IN$, then we can divide the curve through by $k^6$ to give $(a^2 c / k^3)^2 = (ac / k^2)^3 + (-a^2bc/k^4) (ac / k^2)$ so that $x' = ac / k^2 \in \IQ$ and $y' = a^2 c / k^3 \in \IQ$ are the new rescaled points on the curve. Crucially the factor $k^4$ is removed from $D$ so that ultimately we can take $D$ to be fourth-power free.

For Type 2, again let $c - b = a$ be the ABC-triple but now multiply through by $64a^2 c^3$ and then add $(4abc)^2$ to both sides to give $64a^2c^4 - 64a^2bc^3 + 16 a^2b^2c^2= 64a^3 c^3 + 16 a^2 b^2 c^2$. Recognizing the LHS as a perfect square we can set $y = 8ac^2 - 4abc$, $x = 4ac$ and $D = 4abc$ so that the curve is $y^2 = x^3 + D^2$ and moreover $\beta = \frac{y+D}{2D} = \frac{8ac^2}{8abc} = c/b$ as required.
Similarly, for any cube factor, say $k^3 \in \IN$, of $D$, we can divide the curve through by $k^6$ so that $(x',y') = (x/k^2, y/k^3) \in \IQ^2$ is the new point and $D$ is rescaled to $4abc/k^3$.
Hence we can take $D$ to be cube-free.

Finally, for Type 3, we start with $c = a+b$ and multiply through by $a^2c^3$ to give
$(ac^2)^2 = (ac)^3 + a^2bc^3$ so that $(x,y) = (ac,ac^2)$ and $D = a^2bc^3$ and the curve is as needed. Furthermore, $\beta = y^2/D = a^2c^4/(a^2bc^3) = c/b$ as required.
Suppose $D$ had a sixth-power factor $k^6 \in \IN$, we can simply divide by this to render $D$ sixth-power free.

\begin{figure}[!h]
\centerline{
  (1)
  \includegraphics[trim=0mm 0mm 0mm 0mm, clip, width=2in]{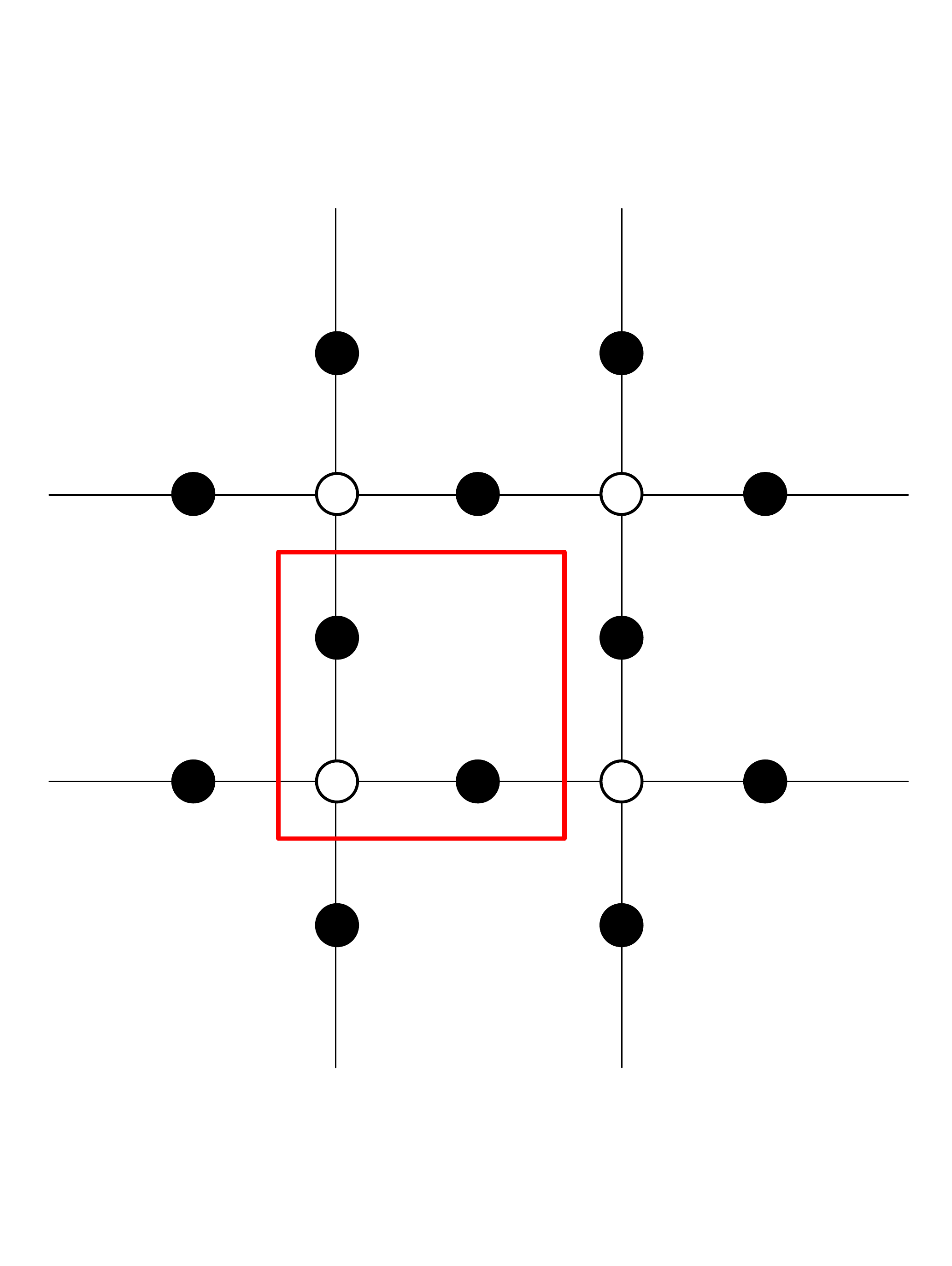}
  (2)
  \includegraphics[trim=0mm 0mm 0mm 0mm, clip, width=2in]{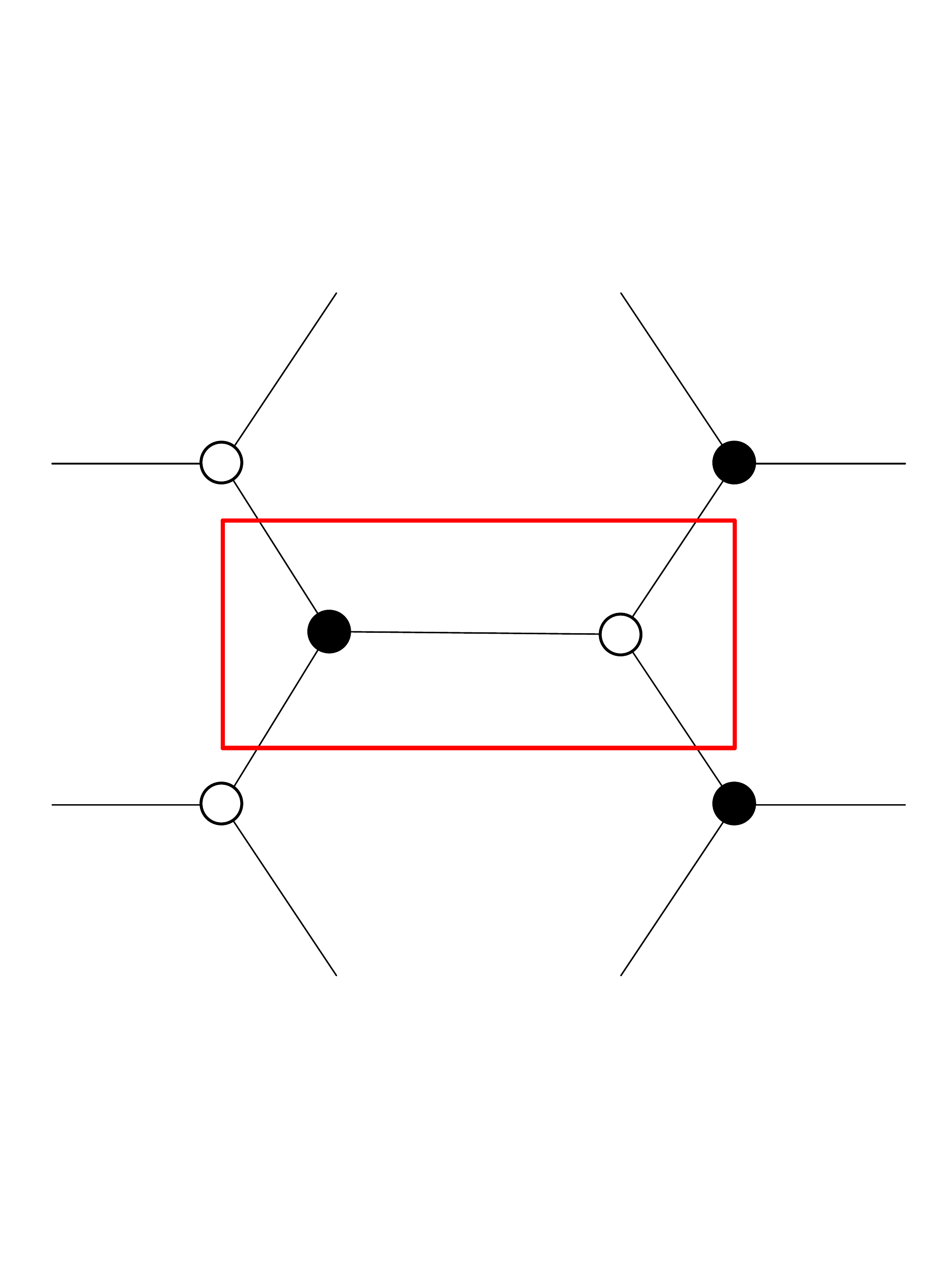}
  (3)
  \includegraphics[trim=0mm 0mm 0mm 0mm, clip, width=2in]{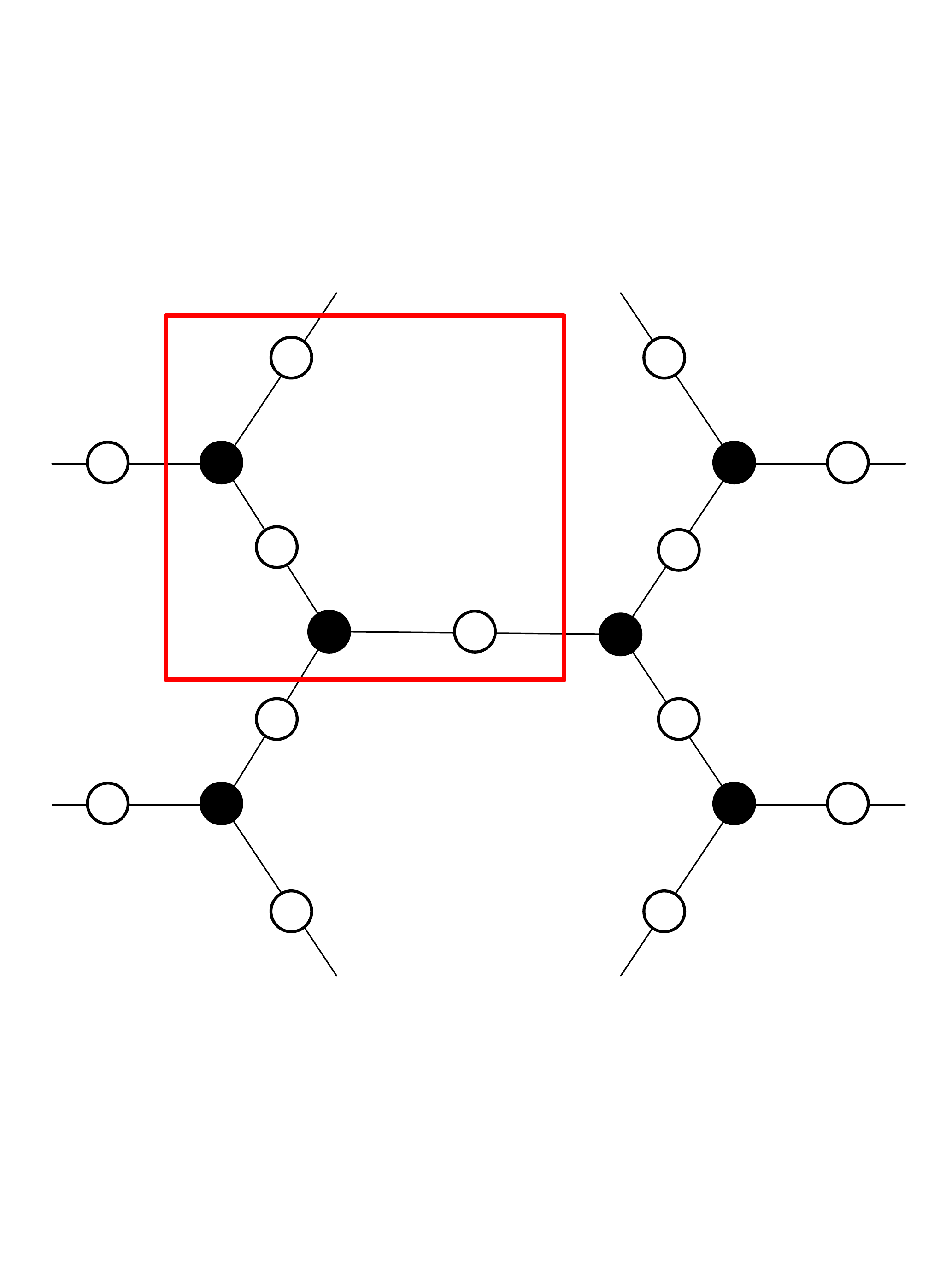}
}
\caption{{\sf {\small
The dessins d'enfants of the 3 types of Belyi pairs in Theorem \ref{thm:KS}. Each is a bipartitite graph on the torus, i.e., a dimer model or brane-tiling. Types (1) and (3) are clean in the sense that one of the colours in completely bi-valent. The fundamental region on the torus is shown in the red box.
The $j$-invariants are, respectively, 1728, 0 and 0.
  }}
\label{f:dessinD}}
\end{figure}

For reference, we draw the dessins d'enfants explicitly in Fig.~\ref{f:dessinD} for each of the three Belyi pairs.
A detailed calculation of the pre-images and ramification structure is presented in Appendix \ref{ap:belyi}.
We note that the $j$-invariants are, respectively, 1728, 0 and 0, which are special values corresponding to elliptic curves with complex multiplication and enhanced symmetries of $\IZ/4\IZ$, $\IZ/6\IZ$ and $\IZ/6\IZ$.

It is expedient to translate between an ABC-triple and points on the elliptic curve. For reasons to be explained in the following section, we will focus on Type 2 Belyi maps.
Obviously, given a rational point $(x,y)$ on the curve $E_D$, we have a unique ABC-triple given by $c/b = (y+D)/(2D)$.
The reverse direction is a little more involved, though readily algorithmic.
We present the detailed algorithm for Type 2 -- which we will see shortly to be central to our discussions -- in Appendix \ref{ap:curve}.

\comment{
Let $(x,y)$ be a rational point on an elliptic curve $E$ and write
\begin{equation}
  (x,y) = (\frac{s}{d^2}, \frac{t}{d^3}) \ ; \quad
  s,t,d \in \IZ \ , \ \ \gcd(s,d) = \gcd(t,d) = 1 \ .
\end{equation}
This is always possible because the Weierstra\ss\ form is cubic in $x$ and quadratic in $y$.
}

Now that we can translate between the triples $(a,b,c)$ and $(x,y,D)$, it is straightforward to phrase the ABC Conjecture purely in terms of the curve.
Since the map from $(x,y,D)$ to $(a,b,c)$ is injective, let us consider a given point $(x,y)$ on $E_D$ for some $D$.
Then, simply reduce $(y+D)/(2D)$ into a minimal fraction, call it $c/b$ and set $a=c-b$. Now, form the quality $q(a,b,c) = \log(c)/\log(\rad(abc))$ as usual.
The ABC Conjecture is the statement that $\limsup$ over the quality for all points $(x,y)$ on all curves $E_D = \{y^2 = x^3 + D^2\}$ is equal to 1.

\section{ABC Conjecture and $\cN=4$ Super-Yang-Mills}\setall
The astute reader might have recognized the dessins in Fig.~\ref{f:dessinD}, especially Type 2.
In this section we will delve into this correspondence by mapping ABC-triples into  $\cN=4$ super-Yang-Mills theory both in the context of field theory and of mirror symmetry in string theory.
We begin by summarizing the relevant constructions from the physics and then study how high quality ABC-triples distribute themselves therein amongst the space of all triples.

\subsection{Quiver Gauge Theory and Elliptic Curves}

\begin{figure}[t]
\begin{center}
  (a)
  $\begin{array}{l}
  \includegraphics[trim=0mm 0mm 0mm 0mm, clip, width=1in]{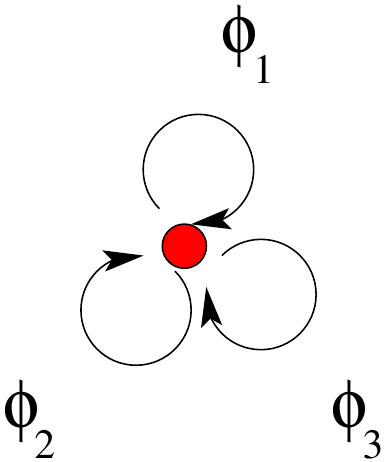}\\
  W = \tr(\phi_1 [\phi_2, \phi_3])
  \end{array}$
  \quad
  (b) $\begin{array}{c} \includegraphics[trim=0mm 0mm 0mm 0mm, clip, width=1.6in]{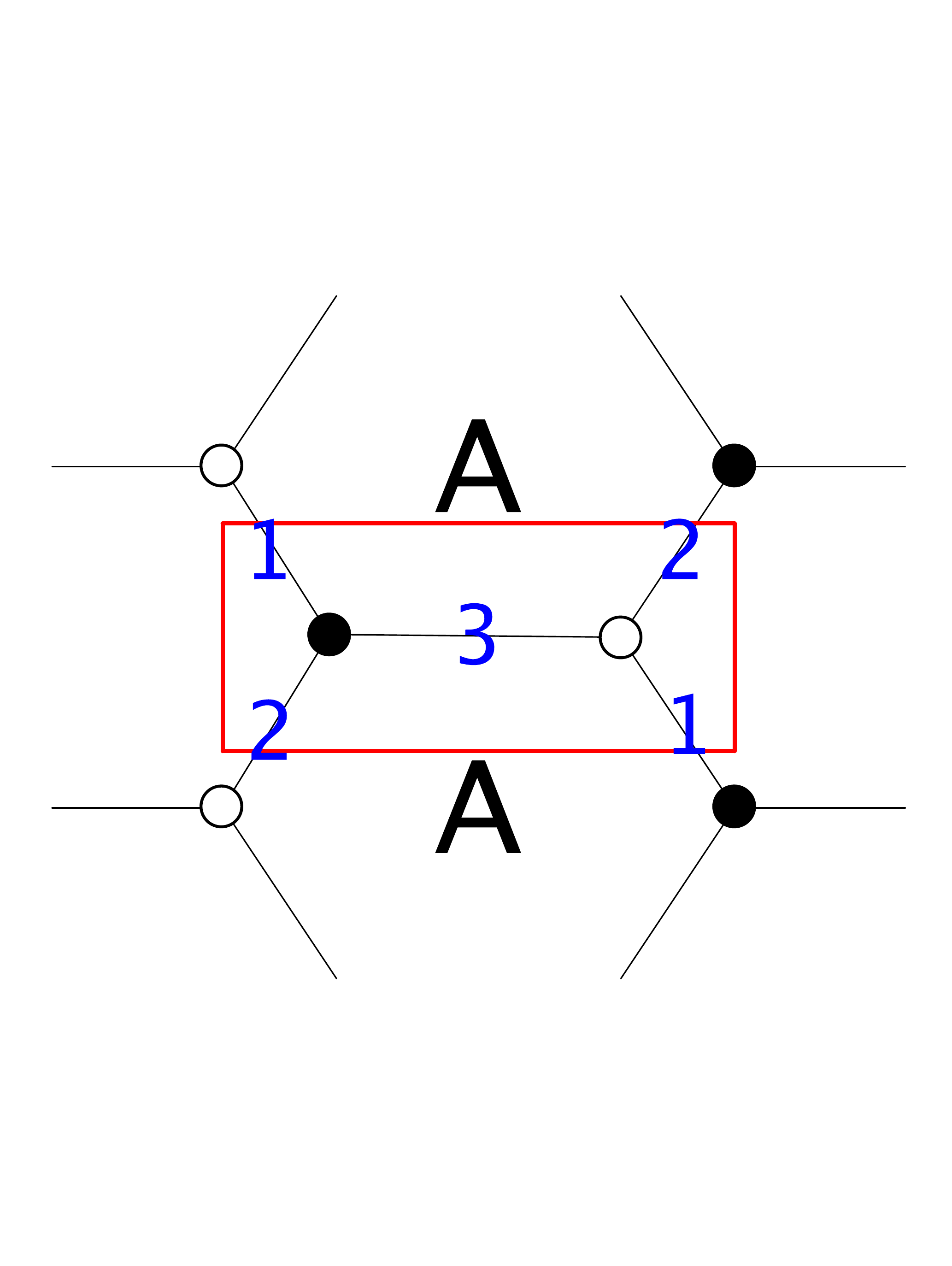} \end{array}$
  (c) $\begin{array}{c} \includegraphics[trim=0mm 0mm 0mm 0mm, clip, width=2in]{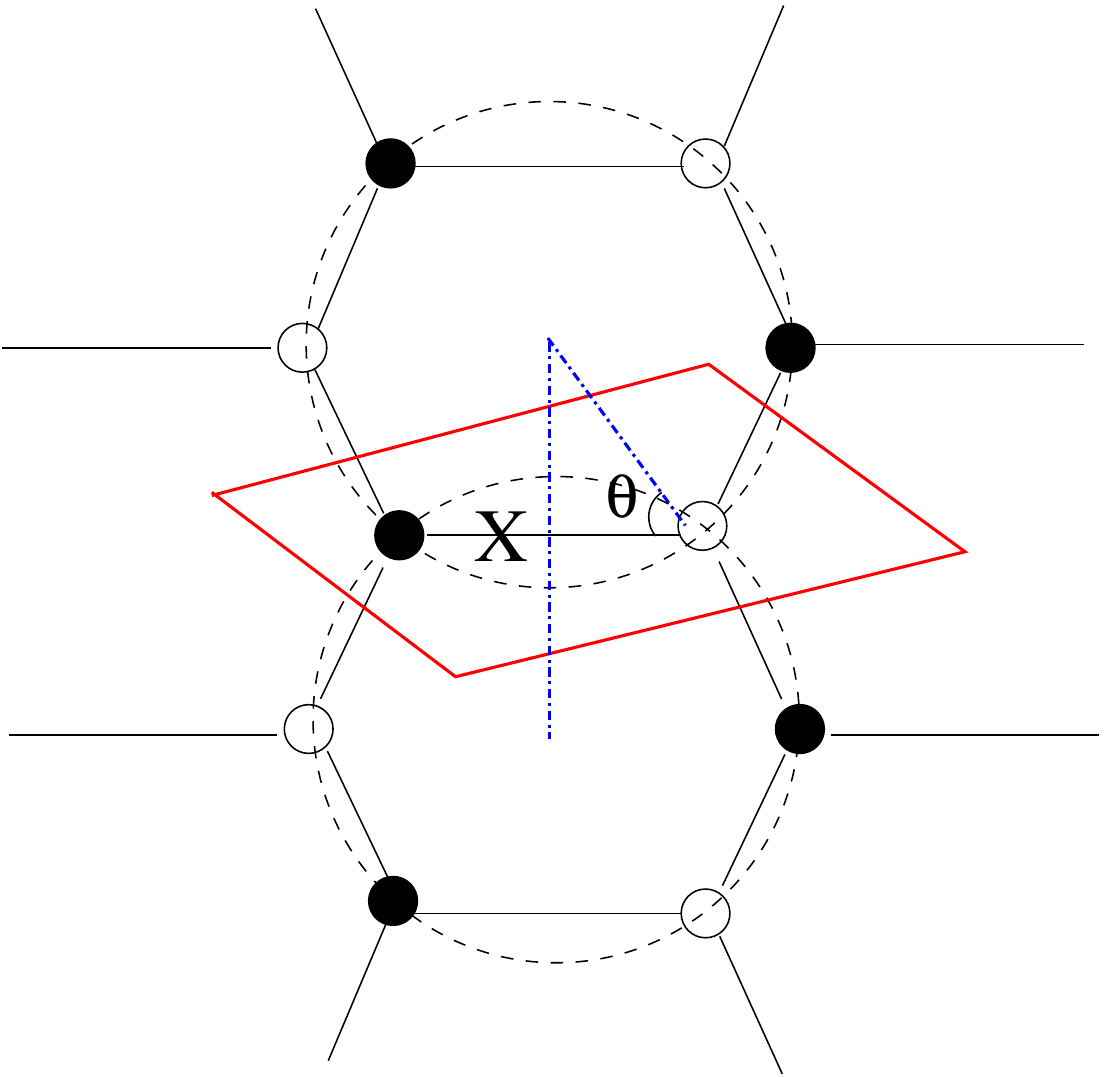} \end{array}$
\end{center}
\caption{
  {\sf {\small (a) Quiver and Dimer for the $\mathcal{N}=4$ super-Yang--Mills theory, corresponding to the toric Calabi--Yau threefold $\mathbb{C}^3$.
      In the quiver, the node corresponds to the $U(N)$ gauge group and the 3 arrows, the 3 adjoint fields $\phi_{i=1,2,3}$.
      (b) In the dimer, the single face, a hexagon  marked by ``A'', corresponds to the gauge group.
      There 3 fields emanate from the trivalent black/white nodes, labeled as $1$, $2$, and $3$.
      The superpotential in terms of these three fields $\phi_{i=1,2,3}$ is the standard
      $W = {\rm Tr}(\phi_1\phi_2\phi_3 - \phi_1\phi_3\phi_2)$, as can be seen going a
      round the white node clockwise and the black node anti-clockwise.
      The diagram is understood to extend doubly periodically and we have drawn, in red, the fundamental region.
      (c) The dimer drawn in isoradial embedding, marked by the dotted unit circles. Here, we have marked one edge as the field $X$, and the angle $\theta$ is related to its $R$-charge as $\theta = \frac{\pi}{2} R(X)$.
\label{f:c3}
}}}
\end{figure}

It had been realized over the years that the gauge theory of a stack of D3-branes probing an affine {\it toric} Calabi-Yau space \cite{Feng:2000mi} is best understood in terms of a bipartite graph tiling a torus, i.e., a {\bf dimer model} or {\bf brane-tiling} on a doubly-periodic plane \cite{Franco:2005rj} (cf.~reviews on the subject in \cite{Yamazaki:2008bt,He:2012js}).
More recently, in an attempt to understand the arithmetic underlying these quiver gauge theories, the brane tiling has been recast into the language of Grothendieck's {\it dessins d'enfants}
\cite{Jejjala:2010vb,Hanany:2011ra,Hanany:2011bs,He:2012xw} where many Belyi pairs have been found explicitly (q.v.~\cite{He:2014jva,Bose:2014lea,He:2015vua} for maps beyond genus 1).

Take the most famous example of a supersymmetric gauge theory: $\cN=4$ super-Yang-Mills (SYM) theory in $3+1$-dimensions with gauge group $U(N)$. There are 3 fields $\phi_{i=1,2,3}$ in the adjoint representation of $U(N)$ with the superpotential $W = \tr(\phi_1 [\phi_2, \phi_3])$. This may be represented as a ``clover'' quiver as shown in part (a) of Fig.~\ref{f:c3}, where the node is the $U(N)$ gauge group and the 3 arrows are the fields; the superpotential is given as an extra piece of data the Jacobian of which gives the relation on the quiver.
The dual graph of this is the bipartite hexagonal tiling of $T^2$. Here, the labeled edges still correspond to the 3 fields and around the black node we proceed counter-clockwise to give the $\phi_1 \phi_2 \phi_3$ term in $W$, while around the white node we proceed clockwise to give the other  $\phi_1 \phi_3 \phi_2$ term in $W$.
The fact we are tiling $T^2$ is enforced by the constraint that $N_g - N_f + N_t = 0$, the Euler number of the torus; here in general $N_g$ is the number of gauge groups (nodes in the quiver), $N_f$, the number of fields and $N_t$, the number of terms in the superpotential. This bipartite graph is shown in part (b) of Fig.~\ref{f:c3}.

Being a bipartite graph on a Riemann surface, we can recast it as a dessin, given by the Belyi pair
\begin{equation}\label{c3}
y^2 = x^3 + 1 \ , \qquad \beta(x,y) = \frac12 ( y + 1 ) \ ,
\end{equation}
whereupon one can check (cf.~Appendix \ref{ap:belyi}) that the passport is
{\tiny $\begin{Bmatrix}
    3\\
    3\\
    3\\
  \end{Bmatrix}$} and the detailed ramification structure is
\begin{equation}\label{c3eg}
\begin{array}{|c|c|c|c|c|c|}
\hline
\mathbb{T}^2: y^2 = x^3 + 1 & \stackrel{\beta=\frac12(1+y)}{\longrightarrow} & \mathbb{P}^1 & \mbox{Local Coordinates on }\mathbb{T}^2 & \mbox{Ramif.~Index}(\beta) \\ \hline\hline
(0,-1) & \stackrel\beta\mapsto & 0 & (x,y) \sim (\epsilon,-1-\frac12\epsilon^3) & 3 \\ \hline
(0,1) & \stackrel\beta\mapsto & 1 & (x,y) \sim (\epsilon,1+\frac12\epsilon^3) & 3 \\ \hline
(\infty,\infty) & \stackrel\beta\mapsto & \infty & (x,y) \sim (\epsilon^{-2},\epsilon^{-3}) & 3 \\ \hline
\end{array}
\end{equation}

\subsubsection{Trivalent Dessin and the Type 2 Belyi Pair}

Examining the three dessins in Fig.~\ref{f:dessinD}, we see that two of them are so-called {\em clean}, which means that all nodes of a particular colour have valency 2.
Whilst this is interesting mathematically, the physical situation is less so: valency 2 means that we have a quadratic term contributing to the superpotential; such terms are mass terms and can be integrated out by their equation of motion. Indeed, customarily, in a supersymmetric quantum field theory, the superpotential consists of terms of order at least cubic -- which is also the case for potentials in non-supersymmetric field theories -- and quadratic terms are absorbed into the kinetic terms.
Now, since Theorem \ref{thm:KS} dictates that {\em all} ABC-triples arise from all three types of curves, it suffices in any event for us to consider only Type 2, which has no valency 2 nodes and is in fact completely trivalent.

As shown in Appendix \ref{ap:belyi}, Type 2 is isogenous over $\IC$ to \eqref{c3} by the scaling $(x,y) \mapsto (D^{2/3} x, D y)$ and we have our familiar $\cN=4$ SYM. Of course, we are working over $\IC$ and do not have this luxury of $D^{2/3}$ and in fact the interesting arithmetic is in the choice of the integer $D$. However, in encoding the gauge theory data, Type 2 for any $D$ is equally valid and the elliptic curve of $j$-invariant 0 will henceforth be central to our discussion.
We have that {\em
  ABC-triples can be mapped to $\cN=4$ super-Yang-Mills theory in $3+1$-dimensions via the Belyi pair
  $\left(y^2 = x^3 + D^2, \ \beta = \frac{y+D}{2D} \right)$ with $D$ cube-free by setting $\beta = c/b$.
}
Henceforth, without ambiguity, we will let $E_D$ denote this Type 2 curve.

Now, examining \eqref{ER}, this highest known quality example was produced on the Type 1 curve. 
We should be able to recast this into our present Belyi pair, which is for Type 2 (cf.¬\cite{KS}).
We have that $\frac{y+D}{2D} = c/b = \frac{23^5}{3^{10} \cdot 109}$, which, upon back-substitution and checking prime powers, yields the solution
\begin{equation}\label{maxq}
  (x,y,D) = (3^{-6}\cdot 2\cdot23^3, \ 3^{-9}\cdot 5\cdot 19\cdot 23^2\cdot 67751,
    \ 3 \cdot 23^2 \cdot 109) \leadsto
  c/b =  \frac{23^5}{3^{10} \cdot 109} \ ,
\end{equation}
as required. Indeed, we see that $D$ is cube-free here.

It is worthwhile to point out that the parity of $b$ is a crucial ingredient.
In the above $(a,b) = (2, 3^{10} \cdot 109)$. Had we chosen instead, as we are free so to do, $b$ to be $2$ and $a$ to be $3^{10} \cdot 109$, then we would have $\frac{y+D}{2D} = c/b = \frac{23^5}{2}$ and the solution
\begin{equation}
  (x,y,D) = (3^4\cdot 23^3\cdot 109, \ 2\cdot 3\cdot 11\cdot 23^2\cdot 109\cdot 292561, \ 3 \cdot 23^2 \cdot 109) \ ,
\end{equation}
which, curiously, is an integer point on the same curve.
Obviously, the quality remains the same.
That the $D$-parameter is the same for both is not a coincidence and we show this in Appendix \ref{ap:curve}.
For any given ABC-triple, we could have drastically different points $(x,y)$ on the same curve $E_D$ corresponding to the permutations of $\{a,b,c\}$ which we discuss in depth in the aforementioned appendix.
Ordering $0<a<b<c$, of course, would give a unique point.

In general, we have that given any ABC-triple, the point of the curve $E_D$ is given by
\begin{equation}\label{xyD}
  (D,x,y) = \left(
  \pm \widetilde{(abc)/2} \ , \quad
  \sqrt[3]{\frac{4acD^2}{b^2}} \ , \quad
  \frac{a + c}{b} D
  \right) \ ,
\end{equation}
where $\widetilde{\prod_{i} p_i^{\alpha_i}} = \prod_{i} p_i^{\alpha_i \bmod 3}$ is the reduction of all prime powers modulo 3.
Note that given any ABC-triple at least one of them is even, so $(abc)/2$ is an integer.
To illustrate, we take the 10 largest known quality ABC-triples to date from \cite{abcpage} and present the $(x,y,D)$ and the point $c/b = \beta = (y+D)/(2D)$ in Table \ref{t:highq}.
In the table, by switching $(a,b)$, we produce 2 different points $(x,y)$ on $E_D$.

\begin{table}[t!!!]
{\scriptsize
  \[
\begin{array}{|c|c|c|c|c|}\hline
c/b & x & y & D & q \\ \hline \hline
 \frac{23^5}{3^{10}\cdot 109} & \frac{2\cdot 23^3}{3^6} & \frac{5\cdot
   19\cdot 23^2\cdot 67751}{3^9} & 3\cdot 23^2\cdot 109 & 1.62991 \\
 \frac{2^{21}\cdot 23}{3^2\cdot 5^6\cdot 7^3} & \frac{2^9\cdot 11^2\cdot
   23}{5^4\cdot 7^2} & \frac{2^2\cdot 11^2\cdot 23\cdot
   48234617}{5^6\cdot 7^3} & 2^2\cdot 3^2\cdot 11^2\cdot 23 & 1.62599 \\
 \frac{2^8\cdot 3^{22}\cdot 5^4}{7\cdot 29^2\cdot 31^8} & \frac{2^4\cdot
   3^8\cdot 5^2\cdot 19\cdot 1307}{31^4} & \frac{2\cdot 3\cdot
   5\cdot 19\cdot 193\cdot 1307\cdot 26015386204481}{31^6} &
   2\cdot 3\cdot 5\cdot 7\cdot 19\cdot 29^2\cdot 31^2\cdot 1307 &
   1.62349 \\
 \frac{2^8\cdot 3^8\cdot 17^3}{5^{11}\cdot 13^2} & \frac{2^4\cdot 3^4\cdot
   17\cdot 283}{5^6} & \frac{2\cdot 3^2\cdot 283\cdot
   8251953691}{5^9} & 2\cdot 3^2\cdot 5^2\cdot 13^2\cdot 283 & 1.58076
   \\
 \frac{5^4\cdot 7}{2\cdot 3^7} & \frac{5^2\cdot 7}{3^4} & \frac{2^2\cdot
   5\cdot 7\cdot 547}{3^6} & 3\cdot 5\cdot 7 & 1.56789 \\
 \frac{2^{11}\cdot 29}{3^{10}} & \frac{2^5\cdot 7\cdot 29}{3^6} &
   \frac{2\cdot 5\cdot 13\cdot 29\cdot 919}{3^9} & 2\cdot 3\cdot
   29 & 1.54708 \\
 \frac{2\cdot 3^3\cdot 5^{23}\cdot 953}{11^{16}\cdot 13^2\cdot 79} &
   \frac{2\cdot 3\cdot 5^9\cdot 7^2\cdot 41^2\cdot 311\cdot
   953}{11^{10}} & \frac{5^2\cdot 7^2\cdot 41^2\cdot 953\cdot
   613474848363909015989}{11^{15}} & 5^2\cdot 7^2\cdot 11\cdot 13^2\cdot
   41^2\cdot 79\cdot 953 & 1.54443 \\
 \frac{11^5\cdot 17\cdot 31^3\cdot 137}{2^9\cdot 3^{17}\cdot 13^2} &
   \frac{5\cdot 11^3\cdot 17\cdot 31\cdot 137}{2^4\cdot 3^{10}} &
   \frac{11^2\cdot 17\cdot 137\cdot 1741\cdot 4969\cdot
   645833}{2^6\cdot 3^{15}} & 2^2\cdot 3^2\cdot 11^2\cdot 13^2\cdot
   17\cdot 137 & 1.53671 \\
 \frac{3^{13}\cdot 11^2\cdot 31}{2^{30}\cdot 5} & \frac{3^5\cdot 11^2\cdot
   13\cdot 19^2\cdot 31}{2^{18}} & \frac{3\cdot 11^2\cdot 13\cdot
   31\cdot 7187\cdot 458599}{2^{27}} & 2^2\cdot 3\cdot 5\cdot
   11^2\cdot 13\cdot 31 & 1.527 \\
 \frac{2^{10}\cdot 5^2\cdot 7^{15}}{17^3\cdot 29\cdot 31^8} & \frac{2^4\cdot
   3^6\cdot 5^2\cdot 7^5\cdot 23\cdot 2269}{17^2\cdot 31^4} &
   \frac{5^2\cdot 23\cdot 487\cdot 2269\cdot
   249605324368789}{17^3\cdot 31^6} & 5^2\cdot 23\cdot 29\cdot 31^2\cdot
   2269 & 1.52216 \\

\hline

\frac{23^5}{3^{10}\cdot 109} & 3^4\cdot 23^3\cdot 109 & 2\cdot 3\cdot
   11\cdot 23^2\cdot 109\cdot 292561 & 3\cdot 23^2\cdot 109 &
   1.62991 \\
 \frac{2^{21}\cdot 23}{3^2\cdot 5^6\cdot 7^3} & 2^9\cdot 3^2\cdot 5^2\cdot
   7\cdot 23 & 2^2\cdot 3^2\cdot 19\cdot 23\cdot 227\cdot 22367 &
   2^2\cdot 3^2\cdot 11^2\cdot 23 & 1.62599 \\
 \frac{2^8\cdot 3^{22}\cdot 5^4}{7\cdot 29^2\cdot 31^8} & 2^4\cdot 3^8\cdot
   5^2\cdot 7\cdot 29^2\cdot 31^4 & 2\cdot 3\cdot 5\cdot 7\cdot
   29^2\cdot 31^2\cdot 41579743\cdot 241510369 & 2\cdot 3\cdot
   5\cdot 7\cdot 19\cdot 29^2\cdot 31^2\cdot 1307 & 1.62349 \\
 \frac{2^8\cdot 3^8\cdot 17^3}{5^{11}\cdot 13^2} & 2^4\cdot 3^4\cdot 5^5\cdot
   13^2\cdot 17 & 2\cdot 3^2\cdot 5^2\cdot 13^2\cdot 2953\cdot 5588861
   & 2\cdot 3^2\cdot 5^2\cdot 13^2\cdot 283 & 1.58076 \\
 \frac{5^4\cdot 7}{2\cdot 3^7} & 2\cdot 3^3\cdot 5^2\cdot 7 & 3\cdot
   5\cdot 7\cdot 13\cdot 673 & 3\cdot 5\cdot 7 & 1.56789 \\
 \frac{2^{11}\cdot 29}{3^{10}} & \frac{2^5\cdot 3^4\cdot 29}{7^2} &
   \frac{2\cdot 3\cdot 29\cdot 83\cdot 1427}{7^3} & 2\cdot
   3\cdot 29 & 1.54708 \\
 \frac{2\cdot 3^3\cdot 5^{23}\cdot 953}{11^{16}\cdot 13^2\cdot 79} &
   \frac{2\cdot 3\cdot 5^9\cdot 11^6\cdot 13^2\cdot 79\cdot
   953}{311^2} & \frac{5^2\cdot 11\cdot 13^2\cdot 23\cdot 79\cdot
   83\cdot 953\cdot 12689\cdot 50651630398961}{311^3} & 5^2\cdot
   7^2\cdot 11\cdot 13^2\cdot 41^2\cdot 79\cdot 953 & 1.54443 \\
 \frac{11^5\cdot 17\cdot 31^3\cdot 137}{2^9\cdot 3^{17}\cdot 13^2} &
   \frac{2^5\cdot 3^7\cdot 11^3\cdot 13^2\cdot 17\cdot 31\cdot 137}{5^2}
   & \frac{2^2\cdot 3^2\cdot 11^2\cdot 13^2\cdot 17\cdot 137\cdot
   16007\cdot 21391\cdot 65269}{5^3} & 2^2\cdot 3^2\cdot 11^2\cdot
   13^2\cdot 17\cdot 137 & 1.53671 \\
 \frac{3^{13}\cdot 11^2\cdot 31}{2^{30}\cdot 5} & \frac{2^{12}\cdot
   3^5\cdot 5\cdot 11^2\cdot 31}{19^4} & \frac{2^2\cdot 3\cdot 5\cdot
   11^2\cdot 31\cdot 67\cdot 173\cdot 271\cdot 3613}{19^6} &
   2^2\cdot 3\cdot 5\cdot 11^2\cdot 13\cdot 31 & 1.527 \\
 \frac{2^{10}\cdot 5^2\cdot 7^{15}}{17^3\cdot 29\cdot 31^8} & \frac{2^4\cdot
   5^2\cdot 7^5\cdot 17\cdot 29\cdot 31^4}{3^{12}} & \frac{5^2\cdot
   29\cdot 31^2\cdot 53\cdot 139\cdot 32992389167371}{3^{18}} &
   5^2\cdot 23\cdot 29\cdot 31^2\cdot 2269 & 1.52216 \\
\hline

\end{array}
\]
}
\caption{
  {\sf {\small
      The higest quality ABC-triples known as of 2016,
      compiled over the decades.
      We find the corresponding points on the type 2 curve $E_D = \{y^2 = x^3 + D^2\}$ using the Belyi map $c/b = \beta := (y+d)/(2D)$. By switching $(a,b)$, we find precisely two points on the curve.
      \label{t:highq}
}}}
\end{table}

\subsection{String Theory Realization}
It should be emphasized that the elliptic curve is {\em not} merely an auxiliary object. Rather, its affine coordinates have precise meaning when embedded into string theory, wherein the entire construction was first engendered \cite{Feng:2000mi,Franco:2005rj}.
Consider a stack of $N$ parallel coincident D5-branes in type IIB superstring theory in $\IR^{1,9}$ with spacetime coordinates $x^{0,1,\ldots9}$, the world volume is $5+1$-dimensional and occupies, say, $x^{0,1,2,3,5,7}$, as is customary.
After compactifying the $x^5$ and $x^7$ directions, we have the stack of D5-branes wrapping $T^2 = S^1 \times S^1$ and the large-scale world-volume theory is precisely that of $\cN=4$ SYM in $3+1$-dimensions.

More generally -- though we do not need it in detail in our present investigations -- we have a {\it fivebrane system} of D5, NS5 and NS5' branes which will break the supersymmetry down to the more phenomenologically relevant $\cN=1$ and the SYM to a quiver gauge theory. The diagramatic representation for occupation of dimensions by the various branes is traditionally taken to be
\begin{equation}
\begin{array}{c|cccccccccc}
  & 0 & 1 & 2 & 3 & 4 & 5 & 6 & 7 & 8 & 9 \\ \hline
  \mbox{D5} & \times & \times & \times & \times & & \times & & \times & & \\
  \mbox{NS5} & \times & \times & \times & \times & \times & \times & & & \\
  \mbox{NS5'} & \times & \times & \times & \times & & & \times & \times & \\
\end{array}
\end{equation}
The NS5-branes placed orthogonally to the D5-brane can be combined into a single (p,q)-fivebrane, occupying the $x^{0,1,2,3}$ and some two-dimensional subspace of $x^{4,5,6,7}$.
This two-dimensional subspace is another Riemann surface $\tilde{\Sigma}$ which turns out to be described by the Newton polynomial of the toric diagram of the vacuum moduli space of the $\cN=1$ gauge theory.

Since we are only dealing with $\cN=4$ SYM, the situation is particularly simple: we do not need any NS5 branes and the toroidal compactification along the $x^{5,7}$ furnishes the $T^2$.
However, in the dessin language, our $T^2$ is an elliptic curve endowed with complex structure; the study of the various manifestation thereof was the subject of \cite{He:2012xw,Feng:2005gw}.
Algebraically, we can complexify the space-time coordinates as
\begin{equation}\label{xytilde}
  (\tilde{x},\tilde{y})  =
  \left(\exp(2\pi\frac{x^4 + i x^5}{R_1}) \ , \quad \exp(2\pi\frac{x^6 + i x^7}{R_2})\right)
\end{equation}
so that the $x^{5,7}$ directions are periodic with radii $R_1$ and $R_2$ respectively.
We have marked the tildes in order to distinguish from the affine coordinates of the elliptic curve.
Indeed, the toric diagram of $\IC^3$ can be represented as the two-dimensional (the Calabi-Yau condition forces the toric diagram to be planar) polygon with vertices $\cT = \{(0,0), (0,1), (1,0)\}$ and the Newton polynomial is thus
\begin{equation}\label{newtonP}
  P(\tilde{x}, \tilde{y}) = \sum\limits_{(p_i,q_i) \in \cT} a_i\,\tilde{x}^{p_i}\,\tilde{y}^{q_i}
  = a_0 + a_1 \tilde{x} + a_2 \tilde{y} 
\end{equation}
for some arbitrary complex coefficients $a_{0,1,2}$.
This is the explicit algebraic equation for $\tilde{\Sigma}$, which in this case happens to simply be the hyperplane, i.e., genus 0 (one sees that the genus of $\Sigma$ and $\tilde{\Sigma}$ are different in general).
Furthermore, the alga \cite{Feng:2005gw} (or co-am{\oe}ba) projection $(\tilde{x},\tilde{y}) \mapsto (\arg\tilde{x}, \arg\tilde{y})$ retrieves the compact $x^{5,7}$ directions and thus topologically renders the dimer.

\subsubsection{Mirror Symmetry}
Suppose we performed T-duality along the two circles in the $x^{5,7}$ directions of the $T^2$.
This maps the D5-branes to D3-branes and the NS5-branes to pure geometry, in fact, to an affine toric Calabi-Yau threefold $\cM$. The diagram here is
\begin{equation}
  \begin{array}{c|cccccccccc}
    & 0 & 1 & 2 & 3 & 4 & 5 & 6 & 7 & 8 & 9 \\ \hline
    \mbox{D3} & \times & \times & \times & \times & &  & &  & & \\
    \mbox{CY3} & & & & & \times & \times & \times & \times & \times & \times\\
  \end{array}
\end{equation}
and we have the situation of D3-branes placed transversely -- thereby probing the Calabi-Yau threefold.
When the CY3 is trivially $\IC^3$, the $3+1$-dimensional world-volume theory of the stack of $N$ D3-branes is precisely $\cN=4$ SYM.

From this scenario we can perform twice T-duality to return to our fivebrane system but can go one step further and use the celebrated ``thrice-T-duality = mirror symmetry'' paradigm of \cite{Strominger:1996it}. This third $S^1$-direction we will see shortly and what we have now is a stack of D6-branes in type IIA string theory wrapping special Lagrangian 3-cycles of the (local) mirror Calabi-Yau threefold, the  affine equations of which are given as
\begin{equation}
  \{
  u\,v=z ~, \quad
  P(\tilde{x},\,\tilde{y})=z 
  \}
  \subset \IC[u,v,z,\tilde{x},\tilde{y}] \ ,
\end{equation}
where $P$ is the Newton polynomial defined in \eqref{newtonP}.

This mirror threefold $\cW$ can be seen to be a double fibration over $\mathbb{C}$:
(1) the equation $P(\tilde{x},\,\tilde{y})=z$ defines, for each point $z$, the Riemann surface $\tilde{\Sigma}_z$;
(2) the other fibration contains an $S^1$ corresponding to $\{u,\,v\}\rightarrow \{e^{i\,\theta}\,u,\,e^{-i\,\theta}\,v\}$, $U(1)$ which collapses at $z=0$.
This $S^1$ is the third circle along which we perform our last T-duality to map from the fivebrane system to D6-branes.

The surface $\tilde{\Sigma}_z$ develops singularities at some critical points $z_* = z^{cr}_i$ where $\partial_{\tilde{x}} P = \partial_{\tilde{y}} P = 0$ and where a 1-cycle of $\tilde{\Sigma}_z$ pinches off.
Hence, over the segment on the $z$-plane joining $z=0$ and $z_i^{cr}$ there is a $U(1)^2$ coming from the two 1-cycles of $\tilde{\Sigma}_z$, which is pinching off at the ends.
This is topologically an $S^3$ as a $U(1)^2$-fibration over a line segment, and the D6-branes are wrapped thereupon.
There will be one $S^3$ for each critical point of $\tilde{\Sigma}_z$, and all these 3-spheres meet at the origin $z=0$.
The fibre there, $\tilde{\Sigma}_0$, is nothing but the thickened $(p,\,q)$-web associated to the original CY3 which the D3-brane probes, or, equivalently, the $(p,\,q)$-web diagram is the spine of the am{\oe}ba projection of $\tilde{\Sigma}_0$.

\begin{figure}[h!]
(a)\includegraphics[trim=0mm 0mm 0mm 0mm, clip, width=4in]{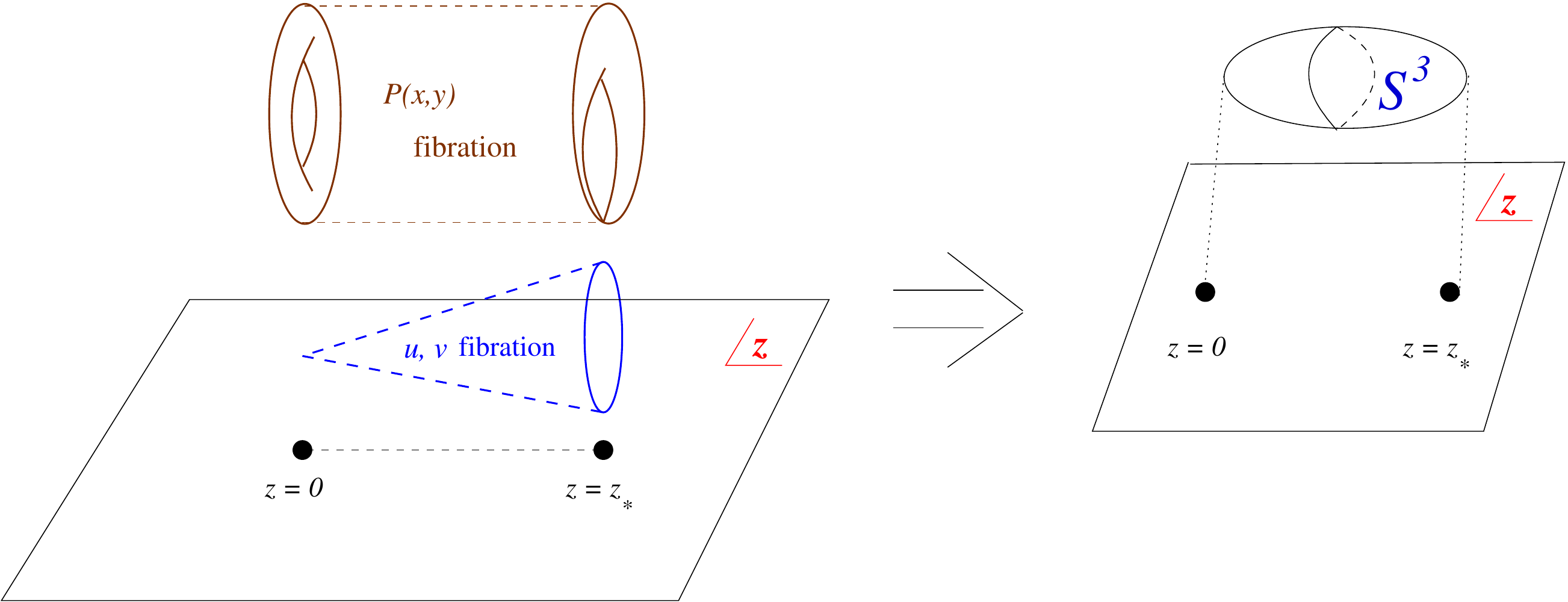}
(b)\includegraphics[trim=0mm 0mm 0mm 0mm, clip, width=2in]{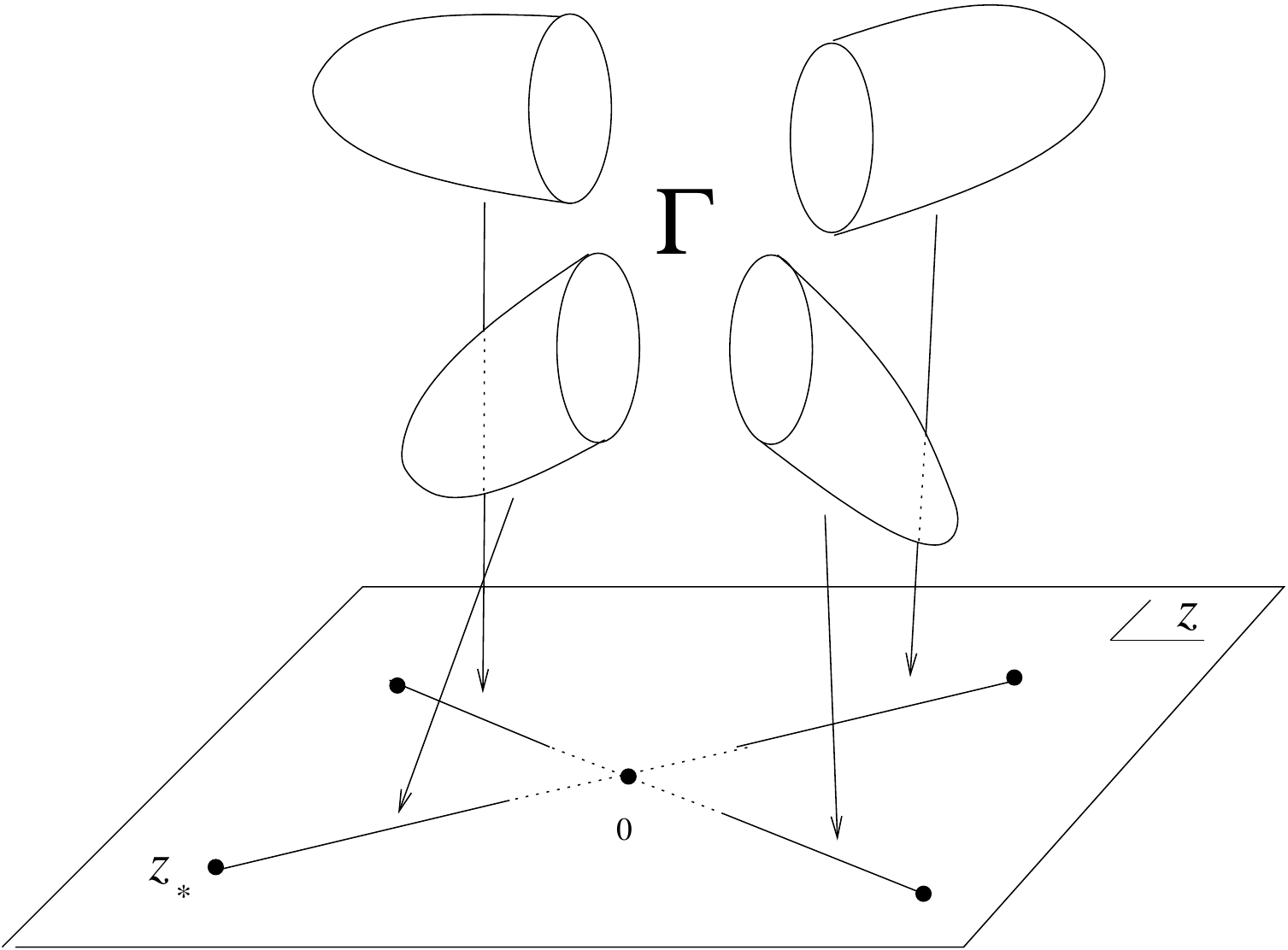}
\caption{{\sf (a) The mirror of the Calabi--Yau threefold as a double fibration over $\IC[z]$.
(b) The $S^3$ cycles meet at the origin in the $z$-plane on a finite graph, which is the dimer model.
}}
\label{f:mirrorfibre}
\end{figure}

The above type IIA picture is illustrated in part (a) of Figure \ref{f:mirrorfibre}.
The dimer itself is then the intersection of these $S^3$ cycles at the origin of the $z$-plane, as some finite graph $\Gamma$; this is shown in part (b) of the figure.

\subsubsection{Special Lagrangian Fibration}
In summary, the $T^2$ is part of the $T^3$ fibre on which one T-dualizes to give mirror symmetry $\cM \leftrightarrow \cW$.
While precisely how this embedding works in general (for arbitrary quiver gauge theories) -- in identifying the Belyi elliptic curve and the $T^2$ on which dimer lives -- remains a puzzle \cite{He:2012xw}, for our case of $\cN=4$ SYM, luckily, it is well understood.
In any of the above scenarios and interpretations, the Belyi elliptic curve embeds into the directions defined by $x^{4,5,\ldots,9}$, which can be complexified by pair-wise combination as in \eqref{xytilde}.

In the prescription of \cite{Strominger:1996it}, $\cM$ is a special Lagrangian (SLag) 3-cycle $\cL$ fibred over some base.
Recall that SLag means that the volume form $\Omega$ and K\"ahler form $\omega$ on $\cM$ both pull back to $\cL$ such that we have the vanishing conditions $\im(\Omega|_\cL) = \omega|_\cL = 0$.
In particular we have the fibration $\cL \hookrightarrow \cM \stackrel{\pi}{\longrightarrow} \IR^3$ so that along the SLag fibres we can thrice T-dualize to $\cW$.

Topologically and metrically, it is straightforward to identify the $T^2$ though algebraically -- i.e., in terms of the coordinates $(x,y)$ of the Belyi pair -- the affine embedding seemingly has a myriad of choices.
For instance, following \S3 of \cite{He:2012xw}, we let the complex coordinates of the $\IC^3$ of $\cN=4$ SYM be $z_1 = x_4 + i x_5, \ z_2 = x_6 + i x_7, \ z_3 = x_8 + i x_9$ in the D3-brane picture where all directions $x^{4,\ldots,9}$ are non-compactified.
The SLag fibration is defined by $\pi : \IC^3_{z_1,z_2,z_3} \mapsto (\im(z_1 z_2 z_3), \ |z_1|^2 - |z_3|^3, \ |z_2|^2 - |z_3|^2) \in \IR^3$.
Now setting $z_j = r_j e^{i \psi_j}$ for $j=1,2,3$, we have the $T^2$ being described by the angular coordinates $(\psi_1, \psi_2)$, with metric $ds_{T^2}^2=\frac{1}{3}\,\Big[ d\psi_1^2+d\psi_2^2+(d\psi_1+d\psi_2)^2\Big]$, which is easily seen to admit complex structure $\tau = \frac{1}{2}\,\Big(1+i\,\sqrt{3}\Big)$, corresponding to the $j$-invariant $j(\tau) = 0$, as is required by the dessin.

The algebraic model for this $T^2$ on which there is a ramified covering of $\IP^1$ at $(0,1,\infty)$ is our curve $y^2 = x^3 + D^2$ with $\beta = \frac{y+D}{2D}$.
Now, crucially, instead of having this $T^2$ over $\IC$, we work over $\IQ$ which suffices to capture all ABC-triples. Note that we do not have a choice on the form of the curve nor can we arbitrarily vary the complex structure: the dessin is a rigid object and even an infinitesimal variation will lead to a dramatically different bipartite graph.
We summarize the above discussions into the following.
\begin{proposition}
  By embedding $\cN=4$ SYM into string theory, we distinguish the elliptic curve $E_D := \{y^2 = x^3 + D^2\}$ of $j$-invariant 0 in two equivalent, mirror-symmetric ways:
  \begin{enumerate}
  \item as a rigid fixing of the the shape of a torus furnished by two compactified directions in a tiling of fivebranes (mirror to D3-branes transverse to $\IC^3$);
  \item as a rigid embedding $T^2$ into the $T^3$-fibre in thrice-T-duality when mapping to D6-branes wrapping 3-cycles in the local mirror of $\IC^3$; 
  \end{enumerate}
  Any ABC-triple can be mapped to a rational point on $E_D$ for $D$ a cube-free integer via the map $c/b = (y+D)/(2D)$.
  Moverover, the ABC conjecture is the statement that $\limsup$ of the quality $\log(c)/\log(\rad(abc))$ over all points $(x,y)$ for all $E_D$ is equal to 1.
\end{proposition}

\subsection{Field Theory Perspective: R-charges and $a$-Maximization}
In the foregoing, the elliptic curve was distinguished by embedding into string theory; we could also understand this purely from the perspective of $\cN=4$ SYM as a field theory.
Now, whilst for general quiver gauge theories, the complex structure in these three points of view -- fivebrane tilings, SLag fibres in mirror symmetry, and the QFT perspective which we now explain -- have yet to be matched and understood \cite{Hanany:2011ra,Hanany:2011bs,He:2012xw}, luckily for the maximally supersymmetric and indeed graph-theoretically symmetric case of $\cN=4$ SYM, the three $\tau$ parameters agree and all distinguish the elliptic curve of $j$-invariant 0.

Let us briefly remind ourselves of rudiments of superconformal gauge theories in $3+1$-dimensions, of which $\cN=4$ SYM is the archetypal example.
We have the following conditions:
\begin{enumerate}
\item For conformality we need to impose the vanishing of $\beta$-functions, both for gauge coupling and superpotential couplings.
For any supersymmetric gauge theory with product gauge group, the NSVZ exact $\beta$-function \cite{Novikov:1983uc} for the $A$-th gauge group factor with coupling $g_A$, in terms of the $R$-charges $R_i$ of all fields $X_i$ charged under $A$, is $\beta_A = \frac{3N}{2}(1-\frac{g_A^2N}{8\pi^2})^{-1}
(2 - \sum\limits_{i: i \in \partial F} (1 - R_i) )$.
For a dimer, the sum runs over the sides which bound a face $F$, indicated by $i \in \partial F$, since faces correspond to gauge groups in the dimer and the edges bounding correspond to fields transforming under that group.
Note that the same field can provide two edges for a single face;
this happens when the field transforms under the adjoint representation of the corresponding gauge group as in $\cN=4$ SYM.
The vanishing of $\beta_A$ for each $A$ requires that for each face in the dimer,
\begin{equation}\label{conf1}
\sum_{ i \in \partial F } ( 1- R_i ) = 2 ~.
\end{equation}
\item The vanishing of the superpotential coupling $\beta$-function gives the condition that, for each node $V$ in the dimer, corresponding to a monomial term in the superpotential,
\begin{equation}\label{conf2}
\sum_{ i : V \in \partial ( i ) } R_i = 2 ~.
\end{equation}
The sum is over edges incident on the vertex $V$, denoted as $V\in \partial (i)$.
\item Finally, we need to perform $a$-maximization \cite{Intriligator:2003jj}, subject to the above constraints of conformality.
  One maximizes the trial $a$-function $a := \frac{3}{32}(3 \tr\,R^3 - \tr\,R)$ for a set of trial $R$-charges, where the trace indicates a 
sum over $R$ charges of the fermions (which are one less than those of 
the bosons in the same multiplet).  
For our theories, $\tr\,R = 0$, so we need only maximize
\begin{equation}\label{a}
a ( \{ R_i\} ) = \sum_{ i =1 }^d ( R_i -1 )^3 ~,
\end{equation}
where the sum is over all the $d$ edges.
\end{enumerate}

As described in~\cite{Hanany:2005ss}, it is particularly convenient to draw the dimer in an \textbf{isoradial} embedding, such that all nodes lie on a circle of unit radius centered on each face.
The $R$-charges of a given bifundamental field $X_{ij}$, on the interface between faces $i$ and $j$ in the dimer, can be encoded in the angle $\theta$ subtended between the edge itself and the radius of the circle centered on face $i$ extending to the node where the $X_{ij}$ edge starts.
Therefore, we have an immediate geometrical formula to read off the $R$-charge:
$\theta=\frac{\pi}{2}\,R[X_{ij}]$.

We demonstrate the relevant quantities for our standard example of $\IC^3$ in part (c) of Fig.~\ref{f:c3}.
Thus, since for a planar isoradial embedding, an $n$-sided polygon has internal angles that sum to $(n-2)\,\pi$ while all angles around a node add up to $2\pi$, the conformality conditions~\eqref{conf1} and~\eqref{conf2} are automatically guaranteed, and it remains only to maximize~\eqref{a} in terms of the angles in the dimer, which can readily be achieved with the help of the computer.
For $\cN=4$ SYM, we have for the 3 fields $\phi_{i=1,2,3}$, after $a$-maximization of the isoradial dimer, $R(\phi_i) = 2/3$, giving the extremal central charge $a_{max} = 1/4$ and complex structure $\tau = \exp(\pi i /3)$.
In summary, we have that
\begin{proposition}
  All ABC-triples arise from the rigid elliptic curve obtained from the torus on which the $a$-maximized isoradial dimer model for $\cN=4$ SYM is represented as a superconformal field theory, viz., the curve $y^2 = x^3 + D^2$ of $j$-invariant 0 and with parameter $D^2$.
  The ABC Conjecture is the statement that $\limsup$ of the quality over all points $(x,y)$ on this family of curves is equal to 1.
\end{proposition}

Given our discussions above in mapping the ABC Conjecture to the elliptic curve which encodes $\cN=4$ SYM, the reader may ask how one actually maps to the dimer model/brane tiling itself.
In other words, if the fundamental domain of the torus as shown in Fig.~\ref{f:c3} actually is comprised of complexified space-time coordinates in the stringy picture, or representing R-charges in the field theory, how do the ABC-triples exhibit themselves therein?
Now we know the fundamental domain has $\tau = \exp(\pi i / 3)$ because of the rigidity of our curve. All points $z \in \IC$ inside are, of course, prescribed by the Weierstra\ss\ $\wp$-function via $(x,y) = (\wp(z), \wp'(z))$.
Indeed, if we choose the parity of $b$ or fix an ordering of $(a,b)$, all the maps are one-to-one.

To set notation and numerical factors, we recall that
\begin{equation}\label{wp}
  y^2 = 4x^3 - g_2 x - g_3 \ ; \qquad
  (x,y) = (\wp(z) \ , \ \wp'(z)) \ .
\end{equation}
We point out that the special case of $(g_2,g_3)=(0,1)$ is called {\it equianharmonic} \cite{AS} and is essentially the case in \eqref{c3}.
Here, the precise periods $(2\omega_1, 2\omega_2)$ which bound the parallelogram of the fundamental region (so that $\wp(z)$ is doubly periodic on the lattice $\IZ(2\omega_1) \oplus \IZ (2\omega_2)$) are
\begin{equation}
  (\omega_1, \omega_2) =
  \left(\frac{\Gamma(1/3)^3}{4 \pi}\ , \ \
  \omega_1 \exp( \frac{\pi i}{3} ) \right)
  \simeq (1.52995, 0.764977 + 1.32498 i)
\end{equation}
with the standard Gamma-function. Note that it is conventional to define the half-periods as $\omega_1$ and $\omega_2$, as is apparent in the definition of the Weierstra\ss\ $\wp$-function
\begin{equation}
  \wp(z) = \frac{1}{z^2} + \sum\limits_{\substack{m,n=-\infty \\ (m,n) \ne (0,0)}}^\infty
  \frac{1}{(z - 2\omega_1 m - 2 \omega_2 n)^2} -
  \frac{1}{(2\omega_1 m + 2 \omega_2 n)^2} \ .
\end{equation}
For our elliptic curve $E_D$, we have that (see details in Appendix \ref{ap:wp})
\begin{equation}\label{wpED}
  (x,y) = (4\wp(z) \ , \ 4\wp'(z)) \ ; \qquad
  ( g_2, \ g_3 ) =
  ( 0, \ - \left(\frac{D}{4}\right)^2 ) \ ; \qquad
  \left(\wp^\prime\right)^2 = 4\wp^3 + \left(\frac{D}{4}\right)^2 \ .
\end{equation}

\begin{figure}[!h]
  \centerline{
    \includegraphics[trim=0mm 0mm 0mm 0mm, clip, width=6in]{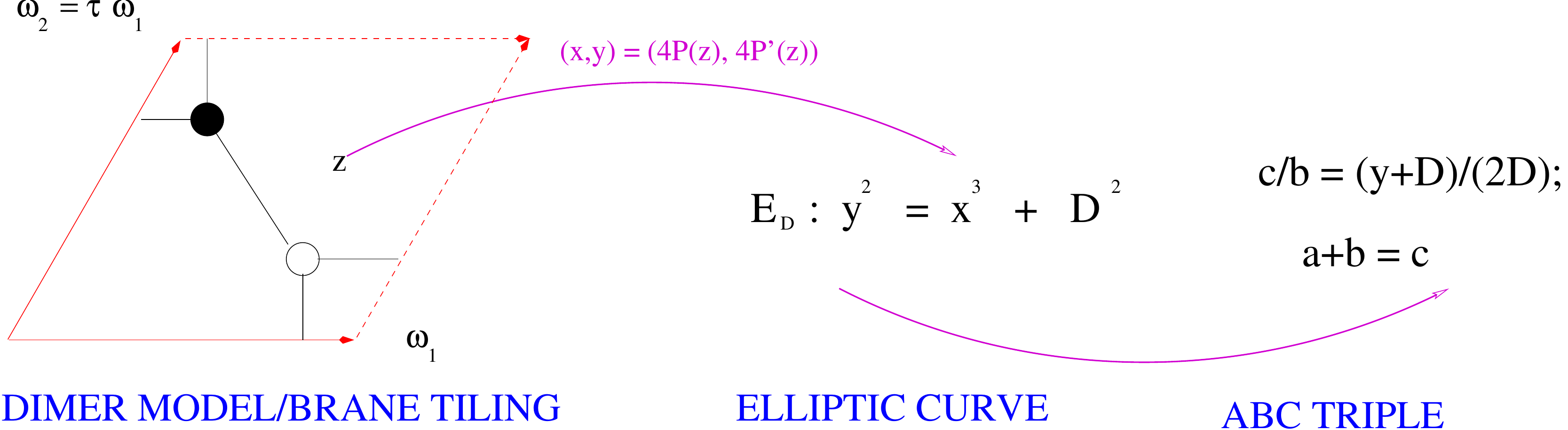}
  }
\caption{{\sf
    {\small Using the Weierstra\ss\ $\wp$-function, we map the points in rigid dimer model/brane tiling (with $j$-invariant 0) for $\cN=4$ SYM theory to ABC-triples. If we fix the parity of $b$ or the ordering of $(a,b)$, the maps shown are 1:1.
  }}
  \label{f:dimer2triple}}
\end{figure}

Graphically, we summarize the discussions of this section in Fig.~\ref{f:dimer2triple}.

A caveat emptor is in order here.
The Weierstra\ss\ $\wp$-function has periods $(2\omega_1, 2\omega_2)$, the ratio of which is $\tau$, which for us is $\exp(\pi i / 3)$ with $j(\tau) = 0$.
Customarily $\omega_1$ is taken to be 1, though we do not have this luxuary here because the parameter $D$ would need to be re-scaled at the cost of $(x,y)$ no longer being rational. 
\comment{
We may take 2 approaches:
\begin{enumerate}
\item
we could sacrifice the rationality of $x$ - since the Belyi map is dependent only on $y$ in any event - by scaling $E_D$ to $ (y/D)^2 = (x / \sqrt[3]{D^2})^3 + 1$. This is the original form of $\cN=4$ in the literature.
Here, $(y + 1)/2 = c/b$ and the ABC-triple would map to $y = \wp(z; \{0,-1\}) = 2c/b-1$ where we have emphasized the $(g_2,g_3)$ dependence of $\wp$ explicitly in the convention of \eqref{wp};
\item we could keep track of how each $D$ value produces a specific pair of half-periods $(\omega_1,\omega_2)$ and hence how the fundamental parallelogram is oriented.
  Indeed since the ratio $\tau$ is fixed to be $\exp(\pi i / 3)$, we have that the length of the two period vectors are the same, and the angle subtended, 60 degrees.
  That is, our fundamental region is a rhombus.
\end{enumerate}
}
We leave a detailed discussion to Appendix \ref{ap:wp} on translating between the half-periods $\omega_i$ of the lattice $\Lambda$ for which our elliptic is isomorphic to $\IC/\Lambda$. We see that we in fact have a scaled and rotated version of the equianharmonic case above and the fundamental domain is a horizontal rhombus (as is indeed required for the isoradial dimer), i.e., 
\begin{equation}\label{dimer2xy}
  \begin{array}{l}
    \includegraphics[trim=0mm 0mm 0mm 0mm, clip, width=3in]{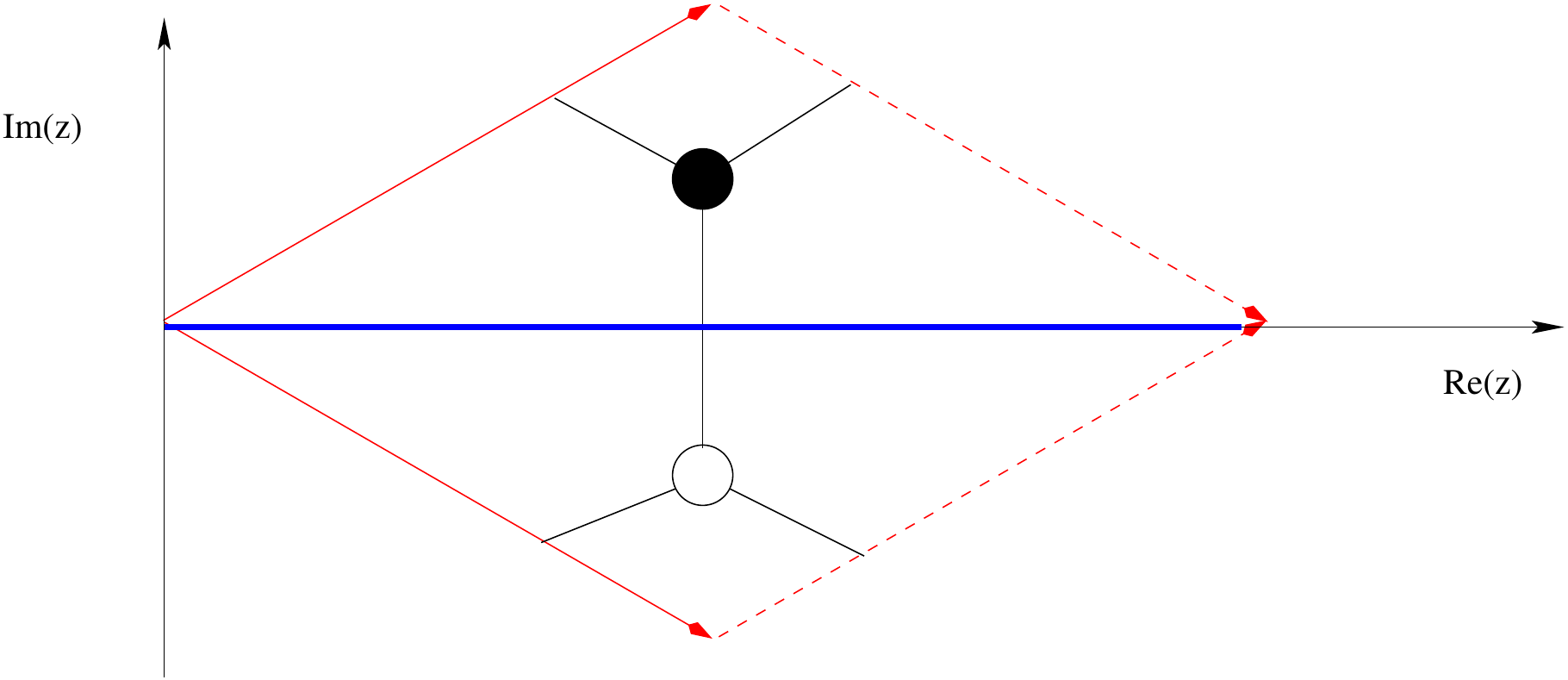}
  \end{array}
  \qquad
  \begin{array}{ll}
    (2\omega_1,2\omega_2) & =
    2 C \sqrt[3]{\frac{4}{D}}
    \left(
    \overline{\zeta_{12}} \ , \ 
    \zeta_{12}
    \right) \\
    & C = \frac{\Gamma\left(\frac{1}{3}\right)^{3}}{4 \pi}
  \end{array}
\end{equation}
where $\zeta_{12}$ is the primitive 12-th root of unity and $C$ is the equianharmonic constant encountered earlier.
We emphasize again that $\omega_{1,2}$ are {\it half}-periods so that the fundamental domains is double the size.

Note that $\omega_2/\omega_1 = \zeta_6 = \exp(2\pi i/6) = \tau$ as required.
This is thus the fixed fundamental region for $E_D$ and the thick blue line on the real axis is mapped to real values of $(x,y)$; this is due to the fact that $\wp(z)$ and $\wp'(z)$ are both meromorphic functions whose Laurent series coefficients are real expressions in $(g_2,g_3)$ which for $E_D$ is equal to $(0, -(D/4)^2)$ by \eqref{WeierstrassVsD}.

In particular, the parameter $D$ does nothing more than to shrink the size of the fundamental domain.
Specifically, the length of the blue line, corresponding to the $(1,1)$-cycle in the torus is
\begin{equation}\label{Lfund}
  L = 2\omega_1 + 2\omega_2 = 
   2 C \sqrt[3]{\frac{4}{D}} (\zeta_{12} + \overline{\zeta_{12}}) =
   2 \sqrt{3} C \sqrt[3]{\frac{4}{D}}
   \ .
\end{equation}

In fact, remembering from \eqref{xytilde} that our two toroidal directions are actually the $x^5$ and $x^7$ directions, we have the interesting fact that
\begin{proposition}
  The parameter $D$ in the elliptic curve $E_D$ dictates the compactification scale for the type IIB string in arriving at the five-brane tiling:
  \[
  R_1 = R_2 \sim 
  \frac{2^{\frac53} 3^{\frac12}  \Gamma\left(\frac{1}{3}\right)^{3}}{4 \pi}
  D^{-\frac13} \ .
  \]
\end{proposition}
Interestingly, $D$ is in general a very large integer, which indeed makes the compactification radius small, in some appropriate units.

Returning to our question about mapping to the fundamental region, recalling from \eqref{xyD} the expression of $x$ in terms of $(a,b,c)$, we have that 
\begin{equation}
\wp(z) = \frac{x}{4} = \sqrt[3]{\frac{acD^2}{(4b)^2}} \ .
\end{equation}
Next, we use the standard fact that the Weierstrass $\wp$-function has an inverse in terms of an elliptic integral on the real axis\footnote{The Weierstra\ss\ function is even, so there is an ambiguity of $\pm$ in the inverse, reflected by whether the $\infty$ in the limit is written above or below. We will choose the {\it positive} half of $\wp^{-1}$.}:
\begin{equation}
  \wp^{-1}(x) = \int_{\frac{x}{4}}^\infty \frac{ds}{\sqrt{4 s^3 - g_2s - g_3}} =
  \int_{\wp(z; \{0, -(D/4)^2\})}^\infty \frac{ds}{\sqrt{4 s^3 + (D/4)^2}} \ ,
\end{equation}
which can be recast into a hypergeometric function to give our final expression mapping ABC-triples onto the blue line in the fundamental domain (which we shall call the {\bf fundamental line}):
\begin{equation}\label{triple2z}
  z = 2 \frac{\, _2F_1\left(\frac{1}{6},\frac{1}{2};\frac{7}{6};-\frac{D^2}{x^3}\right)}{\sqrt{x}} 
     = 2 x^{-\frac12} \, _2F_1\left(\frac{1}{6},\frac{1}{2};\frac{7}{6};-\frac{b^2}{4ac}\right)
     \ , \qquad
  x = \sqrt[3]{\frac{4acD^2}{b^2}} \ .
\end{equation}
Note that since $(x,y)$ are all real here and $x>0$, the expression in the square root in the integral is guaranteed to be positive, whereby validating the hypergeometric representation.
One could, rewriting the hypergeometric function in terms of the associated Legendre functions, express the above as 
$
z = 2^{\frac76} \Gamma(\frac76) (-D)^{- \frac16} x^{-\frac14} 
	P_{-\frac{1}{6}}^{-\frac{1}{6}}\left(\sqrt{\frac{D^2}{x^3}+1}\right)
$.

For reference, combining the definition  $\,_2F_1 (\alpha, \beta; \gamma; z) = \frac{\Gamma(\gamma)}{\Gamma(\alpha)\Gamma(\beta)} \sum\limits_{n=0}^\infty\frac{\Gamma(n+\alpha)\Gamma(n+\beta)}{\Gamma(n+\gamma)\ n!}z^n$ of the hypergeometric function and the simplification of the Gamma function at half-integral arguments: $\Gamma\left(\tfrac{1}{2}+n\right) = {(2n)! \over 4^n n!} \sqrt{\pi}$, the series expansion for $z$ in \eqref{triple2z} can be written succinctly as
\begin{equation}
z = 2 \left( \frac{4acD^2}{b^2} \right)^{-\frac32}  \sum\limits_{n=0}^\infty \frac{(2 n)!}{4^n (6 n+1) (n!)^2} \left( -\frac{b^2}{4ac} \right)^n \ .
\end{equation}

Let us comment on the extent of coverage of the blue fundamental line in \eqref{dimer2xy} by our hypergeometric function.
Recalling that $_2F_1\left(\frac{1}{6},\frac{1}{2};\frac{7}{6};-\frac{D^2}{x^3}\right)$ grows logarithmically for $x > 0$, we have that $z$ in \eqref{triple2z} decreases monotonically for positive $x$ with the maximum attained at
\begin{equation}\label{Ltilde}
\tilde{L} = \lim\limits_{x \rightarrow 0} 2 \frac{\, _2F_1\left(\frac{1}{6},\frac{1}{2};\frac{7}{6};-\frac{D^2}{x^3}\right)}{\sqrt{x}}  = 
\frac{2 \Gamma \left(\frac{1}{3}\right) \Gamma \left(\frac{7}{6}\right)}{\sqrt{\pi } \sqrt[3]{D}} \ .
\end{equation}
Comparing with the length of the fundamental region in \eqref{Lfund}, we see the ratio $\rho = \tilde{L} / L$ nicely reduces to
\begin{equation}\label{ratioL}
\rho = \tilde{L} / L = \frac{2 \sqrt[3]{2} \sqrt{\frac{\pi }{3}} \Gamma \left(\frac{7}{6}\right)}{\Gamma \left(\frac{1}{3}\right)^2}
= \frac13 \ .
\end{equation}
That {\it not} the entire fundamental line is covered should not surprise us: our convention of taking $a,b,c > 0$ makes $x > 0$ on the elliptic curve. Indeed, in general $x,y$ could be negative on $E_D$ and the negative values can be obtained by the group law.
Interestingly, we cover one third of the fundamental line, reflecting the $\IZ_3$ torsion in the Mordell-Weil group.

\section{Points of the Elliptic Curve and Distributions}
\begin{figure}[!h]
  \centerline{
    \begin{tabular}{l}
      (a)
      \includegraphics[trim=0mm 0mm 0mm 0mm, clip, width=2in]{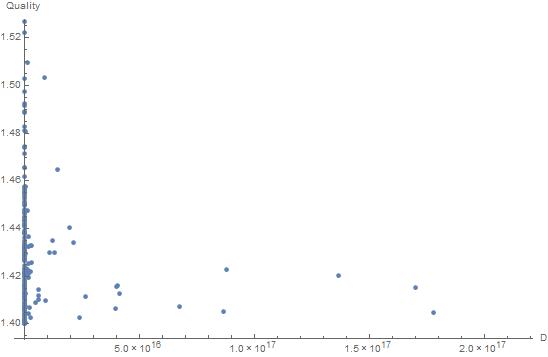}\\
      (b)
      \includegraphics[trim=0mm 0mm 0mm 0mm, clip, width=2in]{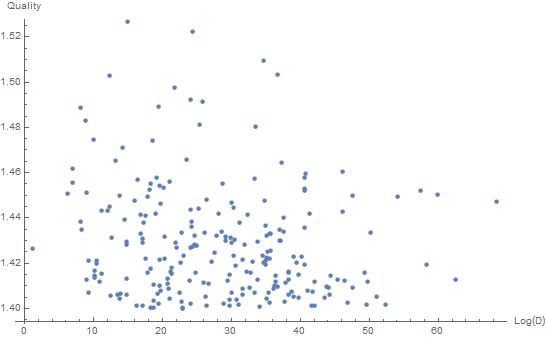}
    \end{tabular}
    (c)$\begin{array}{c}
      \includegraphics[trim=0mm 0mm 0mm 0mm, clip, width=4in]{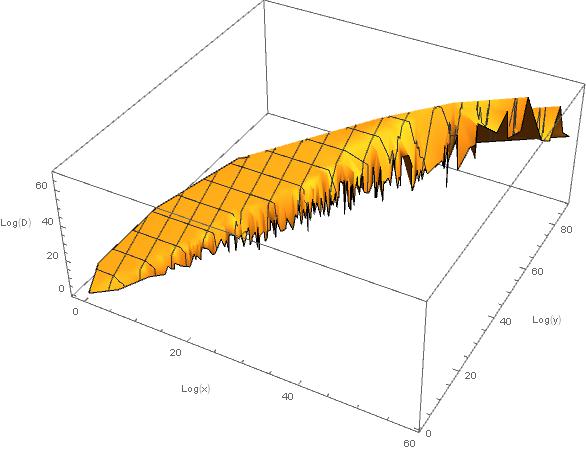}
      \end{array}$
}
\caption{{\sf
    {\small
      (a)
      A scatter plot of the $D$-parameter versus quality for the highest known quality examples of the ABC-triples;
      (b) The same as (a), but with log-axis for $D$;
      (c)
      A 3D-scatter log-plot of the $(x,y,D)$ values for these highest known quality cases.
  }}
\label{f:xyD}}
\end{figure}

\setall
Having singled out our elliptic curve $E_D$, it is expedient to visualize the distribution of ABC-triples thereon.
In part (a) of Fig.~\ref{f:xyD}, we present the (cube-free integer) $D$-parameter versus the quality for the some 200 highest known quality examples \cite{abcpage}.
We see that there is a concentration on lower $D$ values and some sporadic points for extremely large $D$; for clarity we replot this for $\log(D)$ in part (b).
In part (c) of the same figure we draw the scatter plot of the logarithm of the $(x,y,D)$ values for these highest known quality triples.

\begin{figure}[h!!!]
\centerline{
  (a)
  \includegraphics[trim=0mm 0mm 0mm 0mm, clip, width=3in]{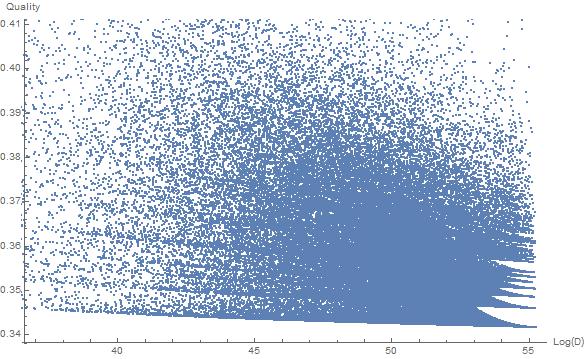}
  (b)
  \includegraphics[trim=0mm 0mm 0mm 0mm, clip, width=3in]{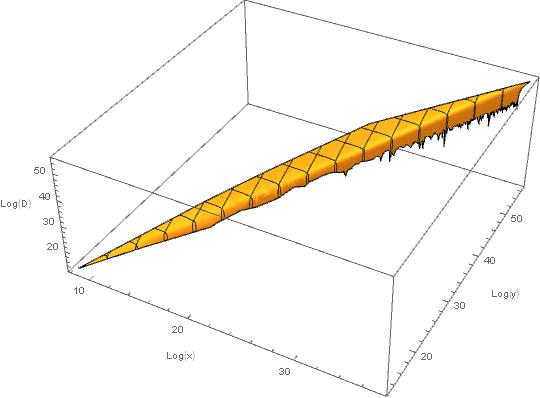}
}
\caption{{\sf
    {\small
      Sampling over $10^5$ random ABC-triples, the corresponding points $(x,y,D)$ on the curve $E_D$ are found.
      (a)
      scatter plot for the the quality of the ABC-triple versus $\log(D)$;
      (b)
      the log scatter plot of all $(x,y,D)$ values.
  }}
\label{f:xyDdist}}
\end{figure}

We can further explore the distribution of the quality and $D$ value for points $(x,y)$ on $E_D$.
We sample over $10^5$ randomly distributed ABC-triples in the interval $[1 , 10^8]$ and find the quality. Then we can determine how the quality is distributed over the corresponding $(x,y)$ values, which we know to be rational. 
\comment{Now, since the Belyi map is dependent only on $y$, distributions of the quality and $D$ will be $x$-independent. In other words, on the curve $E_D$, the pair $(x,y)$ are not independent degrees of freedom.}
In complete analogy to Fig.~\ref{f:xyD}, in part (a) of Fig.~\ref{f:xyDdist}, we present a scatter plot of the quality versus $\log(D)$; we see that the distribution of the quality is intricate and highly unpredictable.
We also see that the quality is generically rather low; over our sample size of $10^5$, we did not even reach a single one exceeding 1/2.
This is why finding ABC triples with high quality, say exceeding 1.4, had been a major undertaking in computer algebra over the past few decades.
Likewise, we present the log-scatter-plot of the $10^5$ $(x,y,D)$-values.
Again, there seems to be an enveloping shape to the possible value.

Now let us re-examine Fig.~\ref{f:dimer2triple} under our present considerations. 
\comment{
  First, since $\wp(z; \{g_2, g_3\})$ has a Laurent expansion in $z$ with coefficients which are real functions in $\{g_2, g_3\}$, which for us is equal to $\{0, -D^2\} \in \IR^2$ and since $(x,y)$ relevant to ABC-triples are clearly real, the points $z$ in the fundamental domain of the brane tiling/dimer model are actually all real.
}
From \eqref{triple2z}, we saw that any ABC-triple is in 1-1 correspondence with a real point on the horizontal (blue) axis of the fundamental domain of the brane tiling, whose size, roughly the compactification scale, is dictated by $D^{-1/3}$.
\comment{
In Fig.~\ref{f:zq}, we study the distribution of quality against these real points $z$. In part (a), we plot the quality versus $\log(z)$ for the highest known triples and in part (b), we do so for our sample of $10^5$ points.
\begin{figure}[h!!!]
\centerline{
  (a)
  \includegraphics[trim=0mm 0mm 0mm 0mm, clip, width=3in]{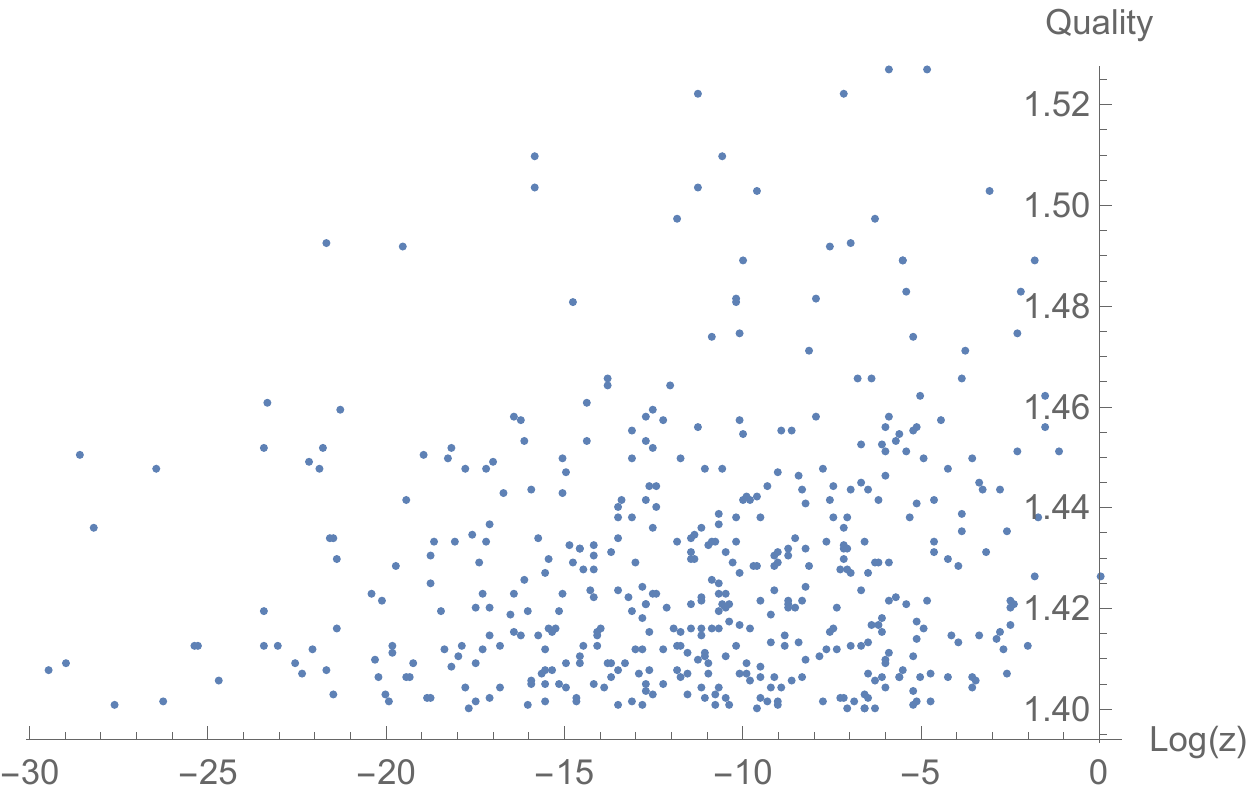}
  (b)
  \includegraphics[trim=0mm 0mm 0mm 0mm, clip, width=3in]{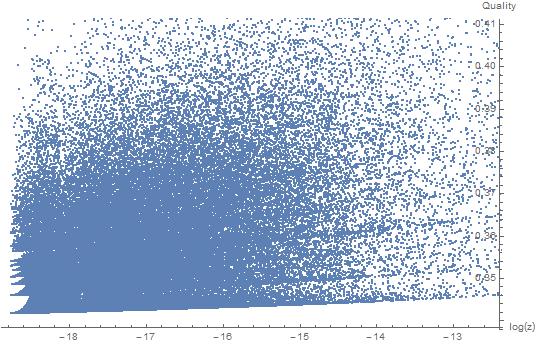}
}
\caption{{\sf
    {\small We map ABC-triples to the fundamental domain;
      they all land on the real axis therein, represented by a real number $z$.
      (a) the quality versus $\log(z)$ for the highest quality cases;
      (b) $q$ versus $\log(z)$ for our sample of $10^5$ triples.
  }}
\label{f:zq}}
\end{figure}
}
Examining the form of \eqref{triple2z}, we see that the prefactor contains $D$ which could be extremely large, as was encountered in Table \ref{t:highq}.
It is illustrative to normalize this factor and indeed the hypergeometric now depends explicitly on the $(a,b,c)$ values, without recourse to the curve $E_D$.
This is also conducive since our toroidal compactification depends on $D$ which could be wildly fluctuating for different ABC-triples and to extract the explicit (the triple still implicitly depends on $(x,y,D)$ via the Belyi map) dependence gives a consistent scale for visualization. 

\begin{figure}[h!!!]
\centerline{
  (a)
  \includegraphics[trim=0mm 0mm 0mm 0mm, clip, width=3in]{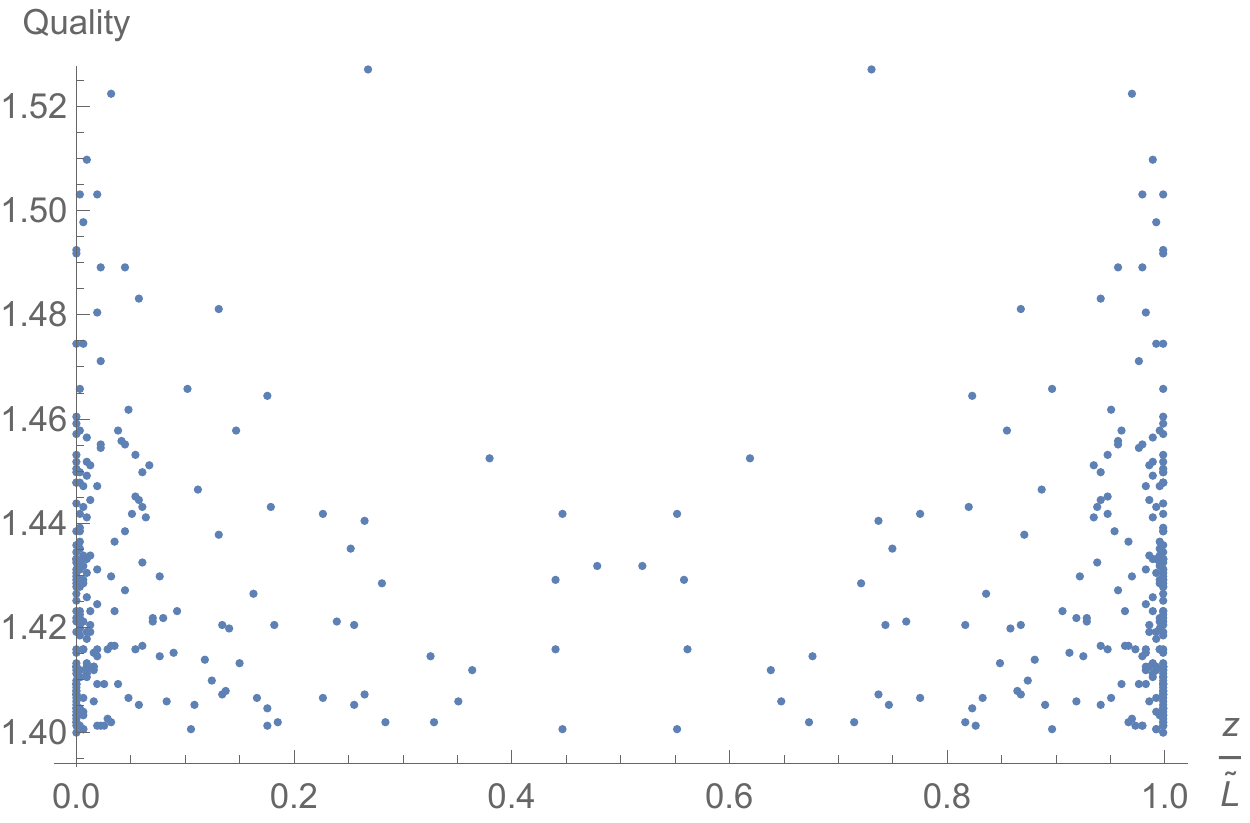}
  (b)
  \includegraphics[trim=0mm 0mm 0mm 0mm, clip, width=3in]{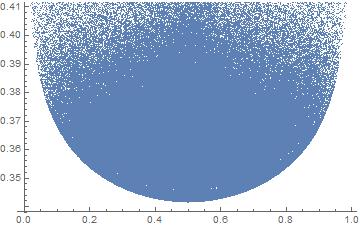}
}
\caption{{\sf
    {\small We map ABC-triples to the fundamental domain;
      they all land on the real axis therein, represented by a real number $z$. We normalize with respect to 1/3 of the length of this line in the fundamental domain due our choice of $(a,b,c)$ being all  positive.
      The quality $q$ on the vertical versus $z / \tilde{L}$ on the horizontal axis for 
      (a) the highest quality cases known;
      (b) a random sample of $10^6$ triples uniformly distributed between 1 and $10^8$.
  }}
\label{f:zqnorm}}
\end{figure}

We discussed earlier in \eqref{ratioL} that the hypergeometric representation of the inverse Weierstrass function covered 2/3 of the fundamental line due our convention of taking $(a,b,c)$-triples to be all positive.
It is therefore illustrative to normalize with respect to this length $\tilde{L}$ and plot the quantity
\begin{equation}
z / \tilde{L} = \frac{\tilde{x}^{\frac16}\sqrt{\pi } \, _2F_1\left(\frac{1}{6},\frac{1}{2};\frac{7}{6};-\tilde{x}\right)}{\Gamma \left(\frac{1}{3}\right) \Gamma \left(\frac{7}{6}\right)} \ , \qquad
\tilde{x} = \frac{b^2}{4 a c} = \frac{D^2}{x^3}
\end{equation}
which is now guaranteed to be between 0 and 1 for $x > 0$.
More precisely, $\frac{z(\tilde{x})}{\tilde{L}}$ is a monotonically increasing function on $\tilde{x} \in \IR_{\ge 0}$ taking the value of 0 at 0 and asymptotically approaching 1 at $\tilde{x} \rightarrow \infty$.
We plot the quality $q$ versus the normalized coordinate $z / \tilde{L}$  in Fig.~\ref{f:zqnorm}, for both the highest quality set as well as a random set of $10^6$ ABC-triples between 1 and $10^8$ (note that we have increased the sample size from the above) and observe some interesting behaviour.

\subsection{Symmetry about 1/2}
The first striking feature is obviously the symmetry about the 1/2-line.
This follows from
\begin{lemma}\label{lem:sym}
  In exchanging $a+b$ and $b+a$, we obtain a reflection on $z(\tilde{x})$ about 1/2, i.e.,
  \[
  \frac{\sqrt{\pi}}{\Gamma \left(\frac{1}{3}\right) \Gamma \left(\frac{7}{6}\right)}
  \left(
  \left( \frac{b^2}{4 a c} \right)^{\frac16} \,
  _2F_1\left(\frac{1}{6},\frac{1}{2};\frac{7}{6};-\frac{b^2}{4 a c} \right)
  +
  \left( \frac{a^2}{4 b c} \right)^{\frac16} \,
  _2F_1\left(\frac{1}{6},\frac{1}{2};\frac{7}{6};-\frac{a^2}{4 b c} \right)
  \right) = 1
  \]
  for $c = a+b$, $a,b,c > 0$.
\end{lemma}
\noindent {\it Proof:}
Adhering to the notation of $\beta = c/b$, we need to show that
\begin{align}
  \nn
  & \left( \frac{1}{4 (\beta-1) \beta} \right)^{\frac16} \,
  _2F_1\left(\frac{1}{6},\frac{1}{2};\frac{7}{6};- \frac{1}{4 (\beta-1) \beta}
  \right)
  +
  \left( \frac{(\beta-1)^2}{4 \beta} \right)^{\frac16} \,
  _2F_1\left(\frac{1}{6},\frac{1}{2};\frac{7}{6};-\frac{(\beta-1)^2}{4 \beta}
  \right) \\
  \label{func}
  & =
  \frac{\Gamma \left(\frac{1}{3}\right) \Gamma \left(\frac{7}{6}\right)}{\sqrt{\pi}} \ ,
  \qquad
  \forall \beta \in (1, \infty) \ .
\end{align}
Defining the LHS of \eqref{func} to be $f(\beta)$, we can readily check that
{\scriptsize
  \[
  \frac{d}{d\beta} f(\beta) =
  \frac{((\beta -1) \beta )^{5/6}-\beta  \left(\sqrt{-(\beta -1) \beta } \sqrt{\beta +\frac{1}{\beta }+2} \sqrt{-(1-2 \beta )^2} \sqrt[6]{\beta +\frac{1}{\beta }-2}+2 \beta  ((\beta -1) \beta )^{5/6}+((\beta -1) \beta )^{5/6}\right)}{3 \sqrt[3]{2} \sqrt{-(1-2 \beta )^2} (-(\beta -1) \beta )^{3/2} (\beta +1)} \ ,
  \]
}
with, importantly, the hypergeometric functions cancelling out.
Moreover, for $\beta > 1$, we have that $f(\beta) = 0$ identically.
Finally, since $\lim\limits_{\beta \rightarrow 1} f(\beta) = \frac{\Gamma \left(\frac{1}{3}\right) \Gamma \left(\frac{7}{6}\right)}{\sqrt{\pi}} $ by the definition of the $_2F_1$, we arrive at our propostion.
\qed

We remark that this curious functional equation for the hypergometric in \eqref{func} holds for all $\beta \in \IC$ with $\re(\beta) \ge 1$ and does not hold for $\re(\beta) < 1$.
In any event, the function equation suffices to show the symmetry about the 1/2 in the normalized fundamental line since in our convention $\IR \ni \beta > 1$.
Other than this trivial symmetry $a \leftrightarrow b$, there are in fact, albeit rare (only 7 below $10^{18}$), completely different ABC-triples with the same quality (cf.~Fig.~7.22 of \cite{palestijn}).

\subsection{Clustering}
Next, we emphasize that, despite appearances, the plots in Fig.~\ref{f:zqnorm} are {\it functions} of the ordinate in terms of the abscissa, since each ABC-triple is uniquely mapped to a value of $z$. Moreover, we see that in part (a) of Fig.~\ref{f:zqnorm}, there is a clustering toward $z / \tilde{L}  = 0$ (and symmetrically, about 1), signifying that, other than sporadic points, when $\frac{b^2}{4ac} \to 0$, high quality triples are manufactured.

Indeed, this reflects the fact that having one of the summands in an ABC-triple being small give high quality examples. We have encountered this earlier in the case of $(a,b) = (2, 3^{10} \cdot 109)$ for the highest known case or in the infinite family $(a,b) = (1, 2^{6n}-1)$ where $q > 1$ for all $n \in \IZ_+$;
we will shortly return to discuss this family in more detail.
For random sampling, we see that the value of $z / \tilde{L}$ being 0 and 1 are all very close to 1 where $b$ is small, so that any relatively high quality can be produced at all.

\begin{figure}[ht!!!]
\centerline{
  (a)
  \includegraphics[trim=0mm 0mm 0mm 0mm, clip, width=3in]{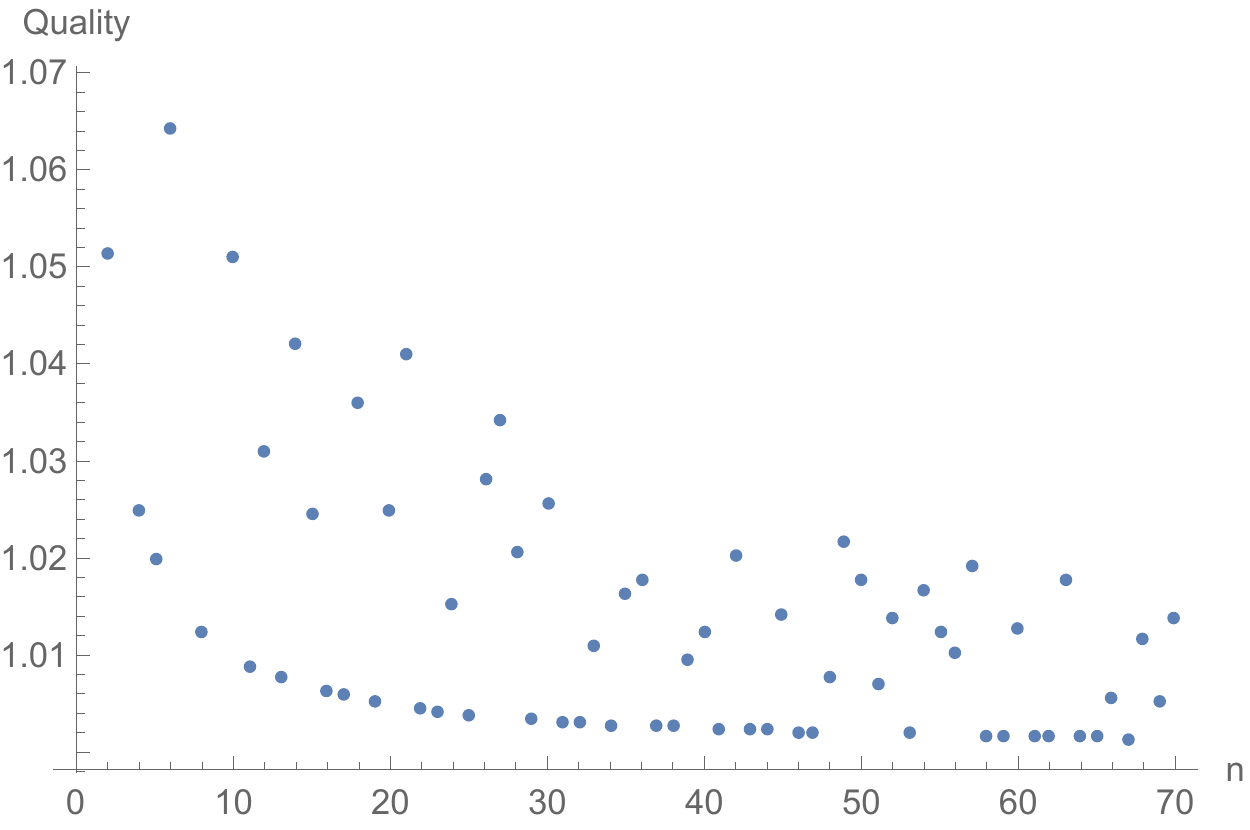}
  (b)
  \includegraphics[trim=0mm 0mm 0mm 0mm, clip, width=3in]{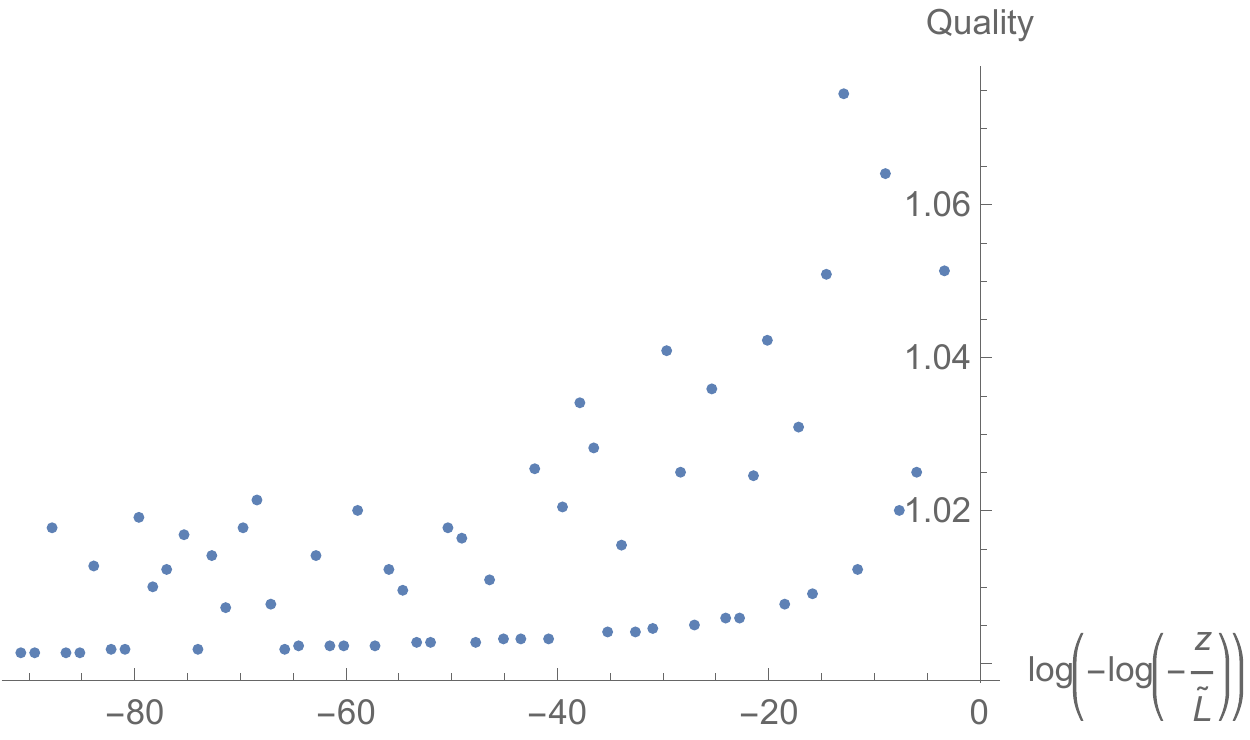}
}
\caption{{\sf
    {\small For the family $(a,b,c) = (1 , 2^{6n}-1, 2^{6n})$ for $n \in \IZ_+$,
      we plot
      (a) the quality $q$ versus $n \in [1, 70]$;
      (b) the quality $q$ versus the double logarithm of the normalized position $\log (- \log (z/\tilde{L}))$ in the fundamental domain.
  }}
\label{f:family}}
\end{figure}

As promised, under this light of infinite families it is beneficial to study this archetypal example:
\begin{equation}
(a,b,c) = (1\ , 2^{6n}-1 \ , 2^{6n}) \ , \quad n \in \IZ_+ \ ,
\end{equation}
all of whose qualities exceed 1, and thus serves as one of the lower envelops to the conjecture (whereby enforcing the need for the $\epsilon$).
In Fig.~\ref{f:family} we show the growth-trend of the quality versus $n$.
In part (a), we simply plot $q$ versus $n$ and in part (b), we plot $q$ versus the double logarithm $\log (- \log (z/\tilde{L}))$ of normalized position in the fundamental region. 
We find a clear patten in the resulting data, with high quality triples concentrated around higher $z/\tilde{L}$ values.
The double-logarithm is needed to distinguish the horizontal scale and gives us an idea of how close these points are to the maximum value of $z/\tilde{L} = 1$.

Incidentally, there is a clear lower bound at the following values of $n$ in our exploration up to 70 (beyond which factorizations in calculating the radical becomes quite prohibitive):
\begin{align}
  \nn
  n = & 2, 4, 5, 8, 11, 13, 16, 17, 19, 22, 23, 25, 29, 31, 32, 34, 37, \\
  & 38, 41, 43, 44, 46, 47, 53, 58, 59, 61, 62, 64, 65, 67 \ldots
\end{align}
Unfortunately, this does not conform to any known sequence of which we are aware. Approximately, a regression fit takes a curve of the form $1 + (10n)^{-1}$.

It should be stressed that there are many examples of infinite families of quality exceeding 1, of which there is an industry (cf.~\S3 of \cite{MM}).
Another example is the sequence $(a,b,c) = (1, c_n-1, c_n)$ with $c_n$ defined recursively as $c_{n+1}  = c_n^4 - 4 c_n^3 + 4 c_n^2$.
The growth rate of the triples here is exponential but one could prove that the quality exceeds $1 + \frac{\log\log c_n}{2 \log c_n} > 1$.

Returning to our clustering phenomenon, we see that in part (b) of Fig.~\ref{f:zqnorm}, sampling randomly does not produce any clustering near 0 or 1, which we know from the abovementioned happens when one of the summands $(a,b)$ is much smaller than the other.
Instead, there seems to be some small clustering near the symmetric 1/2 point.
Incidentally, we should think of part (a) stacked on top of part (b) but since our sampling did not produce any intermediate qualities between 0.4 and 1.4, these data points are missing.
This is another reflection of how difficult it is to produce the high quality ABC-triples, even with 1 million random triples none had quality more than 0.41, where
by making us appreciate indeed the rairty of high quality ones.

\subsection{Lower Bound}
Finally, a seeming and highly non-trivial feature of part (b) is the lower bounding curve.
The lowest point appears to be at the 1/2-symmetry point where the quality is circa 0.34, suggestive of a global infimum.
Let us analyze this in a little more detail.
At $z/\tilde{L} = \frac12$, we actually have an interesting point on the elliptic curve:
\begin{proposition}
  On $E_D$, at the symmetry point $z = z_0 = \frac12 \tilde{L}$, where
  $\tilde{L} := \frac{2 \Gamma \left(\frac{1}{3}\right) \Gamma \left(\frac{7}{6}\right)}{\sqrt{\pi } \sqrt[3]{D}}$,
  we have
  \[
  (x_0, y_0) =
  \left( 4 \wp(z_0; \{0, -D^2/16\}, \ 4 \wp'(z_0; \{0, -D^2/16\} \right)
  = (2 D^{2/3}, 3 D) \ .
  \]
Furthermore, $z_0$ corresponds to the simplest case of $1+1=2$.
\end{proposition}
\noindent {\it Proof: }
It is standard by definition that the Weierstra\ss\ function has double poles at the periods, which in our notation are at
\begin{equation}
  (2 \omega_1, 2\omega_2) =
  \frac{\Gamma \left(\frac{1}{3}\right)^2 \tilde{L}}{2^{\frac73}\sqrt{\pi}\Gamma \left(\frac{7}{6}\right)} \left( \overline{\zeta_{12}} \ , \ \zeta_{12} \right)
\end{equation}
as given in \eqref{dimer2xy}, hence $\wp(z)$ also has a double pole at the lattice point $L = 2\omega_1 + 2\omega_2$ as it does at 0.
Also, $\wp'(z)$ is 0 at the half periods as well as $\omega_1 + \omega_2 = L/2$ on the fundamental line.

The zeros of $\wp(z)$, on the other hand, are notoriously difficult to obtain analytically \cite{EZ}.
Luckily, because we have an explicit representation of the inverse function as the hypergeometric, it suffices to evaluate
$
\lim\limits_{x \to 0}
2 \frac{\, _2F_1\left(\frac{1}{6},\frac{1}{2};\frac{7}{6};-\frac{D^2}{x^3}\right)}{\sqrt{x}}
$
in \eqref{triple2z}, which we recall from \eqref{Ltilde} is precisely $\tilde{L}$.
The other zero of $\wp(z)$ is at $2\tilde{L}$ so that in summary in the fundamental domain of the elliptic curve, $\wp(z)$ has a single pole of order two at the lattice point (say, the origin) as well as two zeros, each of order one, at $\tilde{L}$ and $2\tilde{L}$.
Incidentally, this should be compared with Theorem 1 of \cite{DI} where an explicit evaluation of the integral representation in \cite{EZ} was performed.

However, we need to evaluate $\wp$ and $\wp'$ at the point $\frac12 \tilde{L}$ on the fundamental line, which we know from Lemma \ref{lem:sym} to be a symmetry point and which will not be either a zero or a pole.
Here, the double angle formula for the Weierstra\ss\ function comes to the rescue \cite{AS}:
\begin{equation}
  \wp(2z; \{g_2, g_3\}) =
  \frac{\left(\wp(z; \{g_2, g_3\})^2 + \frac{g_2}{4}\right)^2 + 2 \wp(z; \{g_2, g_3\})}
     {4 \wp(z; \{g_2, g_3\})^3 - g_2 \wp(z; \{g_2, g_3\}) - g_3 } \ .
\end{equation}
Taking $2z = \tilde{L}$, $\{g_2, g_3\} = \{0, -(D/4)^2\}$ and using the aforementioned fact that $\wp(z)$ vanishes there, we can readily solve for our required value of $\wp(z_0)$, as well as the associated value for $\wp'(z_0)$ using the form of $E_D$ to be
\begin{equation}
  \left( \wp(z_0; \{0, -D^2/16\}), \ \wp'(z_0; \{0, -D^2/16\}) \right)
  = \left( \frac{D^{2/3}}{2} \ , \frac34 D \right) \ ,
\end{equation}
implying the first part of our lemma.
We point out that, as always, there are 3 solutions for $x$ corresponding to the 3 cube-roots of unity and 2 solutions of $y$ with $\pm$ sign, but in our convention we take the real positive solution for both.

Let us examine this symmetry point further.
Suppose we had an ABC-triple there, then, even allowing for the $\pm$ sign in $y$, we would have  
\begin{equation}
  \frac{y_0+D}{2D} = \frac{c}{b} \quad \Rightarrow \quad
  c = 2b \mbox{ or } c = -b \ ,
\end{equation}
both of which contradict the very assumptions of the Conjecture that ABC-triples be non-zero and coprime, with
\begin{equation}
1 + 1 = 2 \ , \quad (a,b,c) = (1,1,2) \leadsto q = \log(2)/\log(2) = 1
\end{equation}
being the sole exception because this is the only case where $a = b$.
\qed

How do the actual ABC-triples near this symmetry point behave?
To see this, let us perform a Taylor expansion around $\frac12 \tilde{L}$ in $z$ for the $y$-coordinate of $E_D$, i.e., of $y = 4 \wp'(z, \{0, -(D/4)^2\})$ around
\begin{equation}
  \epsilon := z - z_0 \ , \qquad
  \mbox{ with }
  (x_0, y_0)
  = (2 D^{2/3} \zeta_3^i, \pm 3 D) \ ,
\end{equation}
where we have temporarily restored the full solution with $\zeta_3$ being the primitive cube root of 1 and $i=0,1,2$.
Taking the positive value for $y$, we find that, on defining the combination $\eta := \epsilon D^{1/3}$,
\begin{equation}
  \beta = \frac{c}{b} = \frac{y+D}{2D} =
  2+3 \eta+\frac{9 \eta^2}{2}+\frac{21 \eta^3}{4}+\frac{45 \eta^4}{8}+\frac{45 \eta^5}{8}+\frac{171 \eta^6}{32}+\frac{549 \eta^7}{112}+\cO\left(\eta^8\right)
\end{equation}
Thus written we see the correction terms, in orders of $\epsilon$, of deviation from the 1/2-point as $c$ becomes strictly larger than $b$.
The quality of any ABC-triple is
\begin{equation}
  q = \frac{\log(c)}{\log\rad(abc)} = \frac{\log(b \beta)}{\log\rad((b\beta - b)b^2 \beta)} =
  \frac{\log( 2b + 3 b \eta + \ldots)}{\log \rad(b (2b + 3b \eta + \ldots) (b + 3 b \eta + \ldots )} \ .
\end{equation}
Of course, the radical function does not permit any expansion in $\eta$ and could drastically change in value with respect to any perturbation.

\comment{
At the symmetry point where $\eta = 0$, we have the hypothetical (unattainable unless $a=b=1$) $q_0 = \log(2b) / \log \rad (2 b^3) = \log(2b) / \log \rad (2 b)$.
Any ABC-triple would have the numerator of the quality being clearly larger than that of $q_0$. Furthermore,  \todo{how to show min?}
}

\begin{figure}[ht!!!]
\centerline{
  (a)
  \includegraphics[trim=0mm 0mm 0mm 0mm, clip, width=3in]{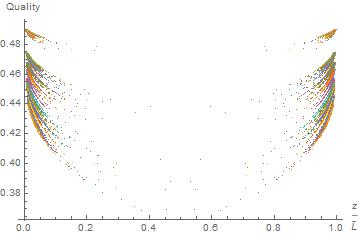}
  (b)
  \includegraphics[trim=0mm 0mm 0mm 0mm, clip, width=3in]{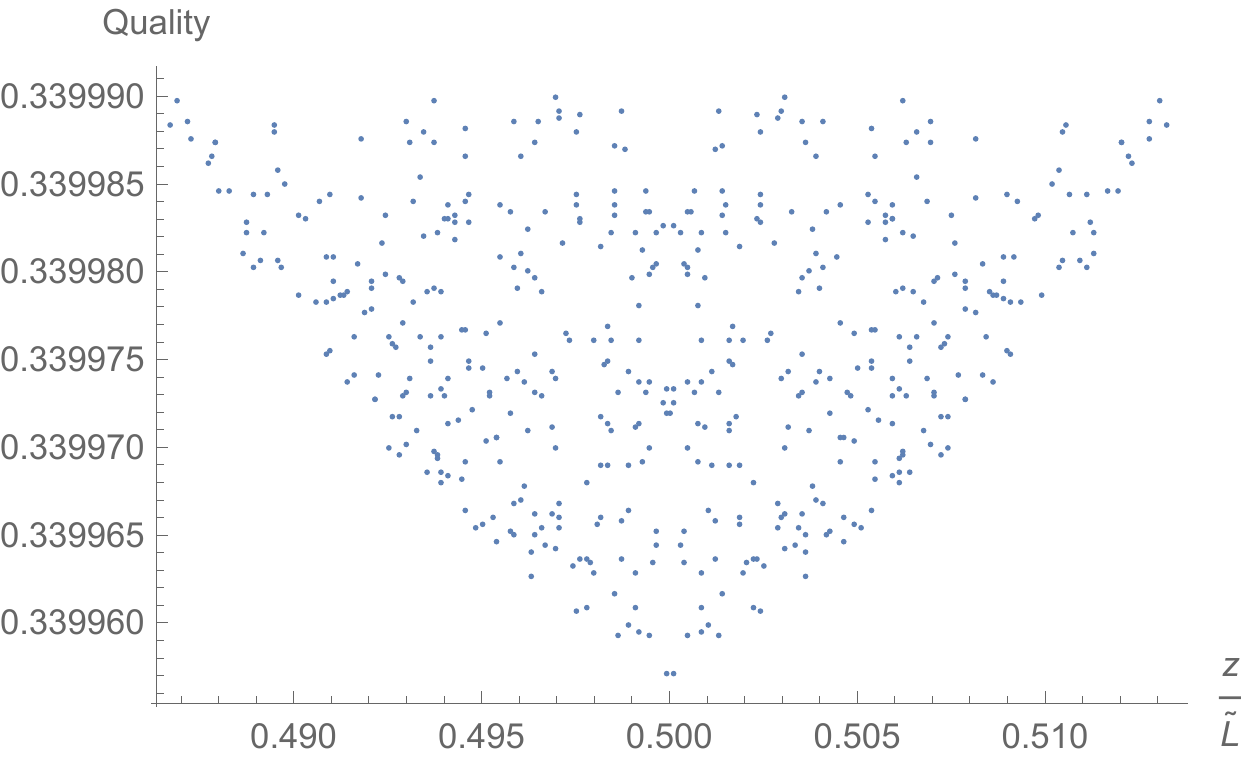}
}
\caption{{\sf
    {\small
      We study some fine structure of the distribution in Fig.~\ref{f:zqnorm} of ABC-triples mapped to the fundamental line by plotting the quality versus the position in the normalized fundamental line for (a) prime pairs with difference $n=2,4,\ldots,100$ (i.e., $(a,b,c) = (p, n, p')$ with $p,p'$ primes); (b) a zoom-in for the lowest quality cases in the random sample of $10^5$ triples between $1$ and $10^{10}$.
  }}
\label{f:finestructure}}
\end{figure}

In any event, there does appear a lower bounding curve in Fig.~\ref{f:zqnorm} (b) with 1/2 being a global infimum.
However, our above proposition shows that the situation is more subtle.
The quality at $\frac12 \tilde{L}$ is equal to 1 and the nearest point to the infimum, within our scan of random ABC-triples between 1 and $10^8$, is achieved by (as well as, of course, its mirror about 1/2)
\begin{equation}
  2 \cdot 3 \cdot 16540169 + 1367 \cdot 73039 = 7 \cdot 127 \cdot 239 \cdot 937
  \ ,
  \quad
  z/\tilde{L} \simeq 0.50072 \ ,
  \quad
  q \simeq 0.341594 \ .
\end{equation}
Interestingly, there are a few points even closer to 1/2, but the qualities are slightly higher: e.g., $98780917 + 3 \cdot 4651 \cdot 7121 = 2 \cdot 5 \cdot 5253 \cdot 6091$ with $z/\tilde{L} \simeq 0.500694$ and $q \simeq 0.341596$ and
$2 \cdot 2153 \cdot 22811 + 98533999 = 3\cdot5 \cdot 13117211$
with $z/\tilde{L} \simeq 0.500374$ and $q \simeq 0.341599$.
Hence, the lower bounding curve, if it exists, has oscillatory behaviour around the apparent envelopping parabola in Fig~\ref{f:zqnorm} and it would be of interest to find it exactly.
By enlargening the sample size to ABC-triples upto $10^{10}$, the above value of $q$ decreases slightly further. Within this range, we find, for instance, about 200 triples out of $10^5$ whose qualities are less than 0.341; we show these in part (b) of Fig.~\ref{f:finestructure}.
It is not obvious whether there is a global minimum for quality.

One might have imagined that the lower bounding curve is constituted of all the {\it twin primes}, since they, being of the form $(a,b,c) = (2, p_k, p_{k+1})$ for some $k \in \IZ_{>1}$ (hence the even prime $2$ is excluded) would have quality
\begin{equation}
  q = \log(q_{k+1}) / \log(\rad(2 p_k p_{k+1})) = \log(q_{k+1}) / \log(2 p_k p_{k+1})
  \ ,
\end{equation}
which tends to make the denominator large as the radical can simply be dropped here.
However, in our random sample, we find this not to be the case.
Nevertheless, we include in part (a) of Fig.~\ref{f:finestructure} the plot for prime-pairs with difference $n = 2, 4, \ldots 100$ ($n=2$ is, of course, the case of twin primes).
We see that each $n$ corresponds to a curve of a parabolic shape, with larger $n$ giving one which is lower.

\subsection{Coupling and S-Duality}
Let us return to the physics.
It is well known (cf.~e.g.,\S4.4.2 of \cite{Yamazaki:2008bt}) that the gauge coupling associated to each gauge group factor in quiver theories of our type is proportional to the area of the torus in the brane tiling.
For SYM, this means that the single coupling constant $\frac{1}{g_{YM}^2} \sim A_F$ where $A_F$ is the area of the rhombic fundamental domain given in \eqref{dimer2xy}.
We easily compute this to be, with $s'$ being some proportionality constant depending on string scale,
\begin{equation}
\frac{1}{g_{YM}^2} = s' |\omega_1| L = 4 s' \sqrt{3} C^2 (4/D)^{2/3} = s D^{-\frac23} \ ,
\end{equation}
where $s = 4^{5/3} \sqrt{3} (4 \pi)^{-2} \Gamma(1/3)^6 s'$.

Importantly, there is an $SL(2;\IZ)$ S-duality action on this {\it complexified} coupling of $\cN=4$ SYM theory, signifying that the action, for $U,V,W,Z\in \IZ$ and $UV - WZ=1$, 
\begin{equation}
  \tau_{YM} \mapsto \frac{U \tau_{YM} + V}{W \tau_{YM} + Z} \ , \quad
  \tau_{YM} := \frac{1}{g_{YM}^2} + i \frac{\theta}{4 \pi} 
\end{equation}
is a symmetry, where $\theta$ is the QCD theta-angle which is a non-geometrical quantity for the brane-tiling. For us, this theta-angle comes from integrating the anti-symmetric B-field over appropriate cycles but whose value can be chosen arbitrarily.
For convenience, we absorb the $4 \pi$ into $\theta$ and consider the $SL(2;\IZ)$ action on $\tau_{YM} = s D^{-\frac23} + i \theta$. Consequently, the transformation, by taking the real part,
\begin{equation}
  s D^{-\frac23} \longrightarrow
  \frac{{D^{2/3}} s (U Z+V W)+D^{4/3} \left(\theta^2 U W+V Z\right)+s^2 U W}
       {2 {D^{2/3}} s W Z+ D^{4/3} \left(\theta^2 W^2+Z^2\right)+s^2 W^2}
\end{equation}
must be a symmetry of the theory {\it for all} values of $\theta \in \IR$ and $U,V,W,Z$ as above.

It is expedient to examine the action by the generators $S = {\tiny \left(\begin{matrix} 1 & 1 \\ 0 & 1 \end{matrix} \right)}$ and $T = {\tiny \left( \begin{matrix} 0 & 1 \\ -1 & 0 \end{matrix} \right)}$ of $SL(2;\IZ)$, corresponding to shift and inversion respectively. These give the actions
\begin{equation}
  s D^{-2/3} \mapsto s D^{-2/3} + 1 \ , \qquad
  s D^{-2/3} \mapsto -\left(s D^{-2/3} + s^{-1} \theta^2 D^{2/3} \right)^{-1} \ .
\end{equation}
Recalling from \eqref{xyD} that $D^2 = \left(\widetilde{(abc)/2}\right)^2$ is in fact a cube-free integer, this means that given an ABC-triple $(a_1,b_,c_1)$, there must exist triples $(a_2,b_,c_2)$ and $(a_3,b_3,c_3)$, such that
\begin{align}
  \nn
  \left(\widetilde{(a_2b_2c_2/2)}\right)^{-2/3}
  & = \left(\widetilde{(a_1b_1c_1/2)}\right)^{-2/3} + s^{-1} \ , \\
  - \left(\widetilde{(a_3b_3c_3/2)}\right)^{2/3}
  & = (s^2 \widetilde{(a_1b_1c_1/2)})^{-2/3} + \theta^2
  (\widetilde{(a_1b_1c_1/2)})^{2/3} \ ,
\end{align}
where as before $\widetilde{n}$ denotes the cube-free part of $n$.
Since we are letting the compactification scale depend on $D$ (and hence on each choice of ABC-triple so that we have a continuum of compactifications as we have the family of elliptic curves $E_D$), we will also allow the string scale $s$ to depend on the individual triples.
This allows for solutions for $(s,\theta)$, for each appropriate trio of ABC-triples.
We note that because of the negative sign in the second transformation, it is expedient to actually use the symmetric version of the ABC-conjecture, i.e., to use triples $a+b+c=0$ so that nagative integers are also allowed.

\subsection{Relations to other Statements}
The ABC Conjecture, has, as mentioned in the introduction, numerous properties regarding elliptic curves.
For instance, there is a classic conjecture of Hall \cite{hall} in the study of Diophantine equations which states that a cube and a square not equal must lie substantially apart.
More precisely, $|y^2 - x^3| > \gamma \sqrt{|x|}$ for some absolute constant $\gamma$.
In modern parlance, a strong version of this conjecture is the following.
Consider the elliptic curve $y^2 = x^3 + k$, then a {\bf Hall-triple} is one for which $(x,y,k)$ are all integers.
A {\it primitive} Hall triple is one for which $\gcd(x^3,y^2)$ is sixth-power free. 
The strong Hall conjecture (cf.~\S12 of \cite{bomb}), equivalent to ABC, is then the statement that for every $\epsilon > 0$, there exists $K_\epsilon$ such that for every primitive Hall triple $(x,y,k)$
\begin{equation}
  |x| \le K_\epsilon |\rad(k)|^{2+\epsilon} \ , \quad
  |y| \le K_\epsilon |\rad(k)|^{3+\epsilon} \ .
\end{equation}
For us, we see that the Hall curve is simply our type 2 and 3 curves.

Let us focus on our type 2 curve $E_D$.
Of the some 200 highest quality triples, we find 47 corresponding to Hall triples on $E_D$. For reference, we tabulate 10 with the highest quality in Table \ref{t:hall}.

\begin{table}[t!!!]
  {\tiny
    \[
  \begin{array}{|c|c|c|c|c|}\hline
    x & y & D & q & \{\frac{x}{\rad(D)^2}, \frac{y}{\rad(D)^3} \\ \hline
 3^4\cdot 23^3\cdot 109 & 2\cdot 3\cdot 11\cdot 23^2\cdot 109\cdot
   292561 & 3\cdot 23^2\cdot 109 & 1.62991 & \{1.9,2.6\} \\
 2^9\cdot 3^2\cdot 5^2\cdot 7\cdot 23 & 2^2\cdot 3^2\cdot 19\cdot
   23\cdot 227\cdot 22367 & 2^2\cdot 3^2\cdot 11^2\cdot 23 & 1.62599 &
   \{8.0,22.8\} \\
 2^4\cdot 3^8\cdot 5^2\cdot 7\cdot 29^2\cdot 31^4 & 2\cdot 3\cdot
   5\cdot 7\cdot 29^2\cdot 31^2\cdot 41579743\cdot 241510369 & 2\cdot
   3\cdot 5\cdot 7\cdot 19\cdot 29^2\cdot 31^2\cdot 1307 & 1.62349 &
   \{0.0,0.0\} \\
 2^4\cdot 3^4\cdot 5^5\cdot 13^2\cdot 17 & 2\cdot 3^2\cdot 5^2\cdot
   13^2\cdot 2953\cdot 5588861 & 2\cdot 3^2\cdot 5^2\cdot 13^2\cdot 283
   & 1.58076 & \{0.96,0.9\} \\
 2\cdot 3^3\cdot 5^2\cdot 7 & 3\cdot 5\cdot 7\cdot 13\cdot 673 &
   3\cdot 5\cdot 7 & 1.56789 & \{0.9,0.8\} \\
 2^4\cdot 5^4\cdot 17\cdot 37^2 & 3^2\cdot 5^2\cdot 7\cdot 37\cdot
   60925103 & 5^2\cdot 37\cdot 239 & 1.50284 & \{0.1,0.04\} \\
 2^8\cdot 3^3\cdot 7^5\cdot 13^2 & 2^2\cdot 3\cdot 7\cdot 11\cdot
   13^2\cdot 30949\cdot 569201 & 2^2\cdot 3\cdot 5^2\cdot 7\cdot
   13^2\cdot 7937 & 1.49762 & \{0.0, 0.0\} \\
 3^2\cdot 5^3\cdot 13^4\cdot 17\cdot 139^2\cdot 151\cdot 4423 & 3^2\cdot
   13\cdot 17\cdot 47\cdot 73\cdot 151\cdot 1399\cdot 4423\cdot
   6691\cdot 438620957 & 2\cdot 3^2\cdot 11\cdot 13\cdot 17\cdot
   151\cdot 4423 & 1.49243 & \{0.1,0.0\} \\
 2^5\cdot 3^6\cdot 7^3\cdot 103\cdot 127\cdot 941^2 & 3\cdot 5\cdot
   7\cdot 31\cdot 127\cdot 673\cdot 941^2\cdot 327823\cdot 349381 &
   3\cdot 7\cdot 73\cdot 127\cdot 941^2 & 1.49159 & \{2.8,4.6\}
   \\
 2^5\cdot 3^3\cdot 5\cdot 13 & 2\cdot 13\cdot 19\cdot 29\cdot 929
 & 2\cdot 11^2\cdot 13 & 1.48887 & \{0.7,0.6\} \\ \hline
  \end{array}\]}
\caption{
  {\sf {\small
      In light of the Hall Conjecture, we tabulate the ratio of $|x|$ and $|y|$ to
      the radical of $k$ on the Hall curve $y^2 = x^3 + k$ for 10 of the highest quality ABC-triples.
      \label{t:hall}
}}}
\end{table}

Another statement, which follows from the uniform ABC conjecture for number fields, is an interesting bound on the actual points on an elliptic curve \cite{KM}.
Adhereing to our notation in Appendix \ref{ap:curve}, let $P = (s/d^2, t/d^3)$ be a point on an elliptic curve $E$ in Weierstra\ss\ form, then there is the conjecture that for all $\epsilon > 0$, there exists a constant $K_\epsilon$ such that
\begin{equation}
  \max \left \{\frac12 \log|s| \ , \ \ \log|d| \right\}
  \le (1 + \epsilon) \log \rad(d) + K_\epsilon \ ,
\end{equation}
for all points $P \in E(\IQ) \backslash \{O\}$.
For our curve and the highest quality ABC-triple thereupon from \eqref{maxq}, we see that $(s,t,d) = (2\cdot23^3, 5\cdot 19\cdot 23^2\cdot 67751, 3^2)$, so that we would be finding a constant $K_\epsilon$ such that 
$\sqrt{2 \cdot23^3} \le K_\epsilon \log(3)^{1 + \epsilon}$.

\section{Conclusions and Outlook}\label{s:conc}\setall
In this paper we have exploited the curious fact that the ABC Conjecture and $\cN=4$ super-Yang-Mills theory, two seemingly drastically unrelated subjects, both localize to exactly the {\it same Belyi pair of the elliptic curve $E_D = \{y^2 = x^3 + D^2\}$ and rational map $\beta = (y+D)/(2D)$.}
It is remarkable that one of the central problems in number theory should conspire to coincide with one of the central objects in theoretical physics.

We have made the correspondence precise by mapping all ABC-triples to the fundamental domain in the brane-tiling for $\cN=4$ SYM; in fact, an injective map is established from ABC-triples to the diagonal $(1,1)$-cycle of the torus representing the real line. We subsequently studied the distribution of the highest known quality triples as well as large random samples of points.
Furthermore, the ABC Conjecture translates to the statement that the $\limsup$ of the quality of the points on this $(1,1)$-cycle should be 1.

Interestingly, the parameter $D$ in the elliptic curve sets the compactification scale in the string embedding of $\cN=4$ SYM: the radii of the toroidal compactification of type IIB superstring which gave rise to the fivebrane tiling system is equal to $\sqrt[6]{\frac{\Gamma\left(\frac{1}{3}\right)^{3}}{16 \pi}} D^{-1/3}$ in Planck units. 

While encouraged by these identifications in realizing a ``physical'' framework for the ABC Conjecture, we have only explored the tip of an iceberg.
Indeed, our usage of $\cN=4$ SYM has been largely kinematical, focusing on the matter content, the superpotential and the brane configuration. We have yet to harness the dynamical: there is a wealth of information ranging from correlation functions, scattering amplitudes and integrability: can these shed light on the ABC Conjecture?

There is a long history of attempting to utilize physical ideas to address profound issues in number theory; the Hilbert-Polya programme for the Riemann Hypothesis, for instance, springs to mind. Here, we have initiated the study of the ABC Conjecture using $\cN=4$ SYM by mapping the triples to brane configurations.
It is our hope that the conjunction of so central a problem in number theory with one so fundamental in physics will initiate beneficial cross-fertilization.

\section*{Acknowledgments}
We are grateful to Sumana Sharma of Trinity College, Cambridge, for tabulating the highest quality ABC-triples and computing the associated $D$ values.
Indebted also are we to Anton Cox, Stefano Cremonesi, Michele del Zotto, and Minhyong Kim for helpful comments and discussions.

YHH acknowledges the Science and Technology
Facilities Council, UK, for grant ST/J00037X/1, the Chinese Ministry of Education, for a ChangJiang
Chair Professorship at NanKai University, and the city of Tian-Jin for a Qian-Ren Award.
YHH is also perpetually indebted to Merton College, Oxford for continuing to provide a quiet
corner of Paradise for musing and contemplations.
ZH is grateful to Chinese National Natural Science Foundation, Grant Number 11501536,
for  visiting London/Oxford, and especially to the ``Workshop on IUT Theory of Shinichi Mochizuki'' at Oxford where this work was initiated.
MP is partially funded by a PROMOS grant issued by the German Academic Exchange Service. 
JR is supported by an AHRC scholarship at Oxford, and is also grateful to Merton College for support.

\appendix
\section{Checking the Belyi Maps}\label{ap:belyi}\setall
We present explicitly the pre-images and ramifications of the three Belyi maps from Theorem \ref{thm:KS}. We will begin with Type 2, which is the most reminiscent of the well-known case of $\cN=4$ super-Yang-Mills.

For Type 2, the map is $\beta = \frac{1}{2D}(y+D)$ so the pre-image of 0 has $y = -D$, whence $(x,y) = \beta^{-1}(0) = (0,-D)$ on the elliptic curve $y^2 = x^3 + D^2$. Choosing local coordinates $(x,y) = (0 + \epsilon, -D + \delta)$ for infinitesimals $(\epsilon, \delta)$, we have that $-2 D \delta = \epsilon^3$. Therefore locally the map is $\beta = \frac{\delta}{2D} \sim -\frac{\epsilon^3}{4D^2}$ and thus the ramification index for this single pre-image of 0 is 3.
Similarly, the pre-image of 1 has $y = D$, i.e., $(x,y) = (0,D)$ is the single pre-image of 1, where local coordinates can be chosen as $(0 + \epsilon, D + \delta)$, so that $\beta \sim \frac{\epsilon^3}{4D^2}$. Hence, the ramification index is also 3 for this single pre-image of 1.
Finally, $(\infty, \infty)$ is the pre-image of $\infty$ where the local coordinates $(\epsilon^{-1}, \delta^{-1})$ can be chosen so that $(x,y) \sim (\epsilon^{-2}, \epsilon^{-3})$.
Hence, the ramification index at $\infty$ is also 3.
In all, the passport for this Type 2 dessin is
{\tiny $\begin{Bmatrix}
3\\
3\\
3\\
\end{Bmatrix}$}, which is easily seen to satisfy \eqref{RR}.
The dessin itself is presented in Part (2) of Fig.~\ref{f:dessinD}.

Interestingly, the above analysis of the pre-images and the ramification indices is independent of $D$.
This would seem to signify that we have a continuous family of inequivalent elliptic curves with Belyi maps, contrary to the notion that dessins are {\it rigid} objects.
Recalling, however, that the $j$-invariant of the elliptic curve is
\begin{equation}\label{j}
  y^2 = 4x^3 - g_2 x - g_3 \qquad \Rightarrow \qquad
  j = 1728 g_3^2 (g_3^2 - 27 g_2^3)^{-1}.
\end{equation}
Then, scaling as necessary, we have that the Type 2 curve has
$(g_2, g_3) = (0, -(D/4)^2)$, whence $j = 0$ for all values of $D$, meaning that this entire family is actually isogenous.
This is unsurprising, since sending $(x,y) \mapsto (D^{2/3} x, D y)$ renders the curve into the familiar $y^2 = x^3 + 1$ and the Belyi map to $\beta = \frac12(y+1)$, our basic case for $\cN=4$ Yang-Mills.

Next, we move onto the rather similar case of Type 3.
Here, the pre-image of 0 has three points $(x,y) = (-\sqrt[3]{D} \omega_3^i, 0)$ for $i=0,1,2$ and $\omega_3$ is the primitive cube root of unity.
Expanding with infinitesimals $(\epsilon, \delta)$, we see that $\delta^2 \sim \epsilon$ so that $\beta = y^2/D \sim \delta^2$ and we have valency 2.
For the pre-images of 1, we have 2 points $(x,y) = (0, \pm \sqrt{|D|})$ so that in local coordinates $\delta \sim \epsilon^3$. Hence the map is $\beta = y^2 / D = (\pm \sqrt{|D|} + \delta)^2/D \sim \delta \sim \epsilon^3$ locally and we have valency 3 for both points.
Finally, the pre-image of $\infty$ is the single point $(\infty, \infty)$ about which we have $(x,y) \sim (\epsilon^{-2}, \epsilon^{-3})$ so the ramification index is 6.
In summary, the passport is
{\tiny
$\begin{Bmatrix}
2,2,2\\
3,3\\
6\\
\end{Bmatrix}$}, which indeed satisfies \eqref{RR}.
Consequently, the dessin is the clean hexaongonal tiling of the torus as shown in Part (3) of Fig.~\ref{f:dessinD}.
The $j$-invariant here is as above: $j = 0$.

At last, we address Type 1.
Here, the pre-image of 0 is the single point $(x,y) = (0,0)$ about which we have the expansion $\delta^2 = (3 +D) \epsilon + \cO(\epsilon^2)$.
Thus, $\beta \sim \epsilon^2 \sim \delta^4$ and $(0,0)$ has valency 4.
The pre-image of 1 consists of the 2 points $(\pm \sqrt{-D}, 0)$ so that thereabout we have $\delta^2 = (\pm \sqrt{-D} + \epsilon)^3 + D (\pm \sqrt{-D} + \epsilon)$, or that $\epsilon \sim \delta^2$. Hence  both these points also have valency 4.
At infinity, the pre-image is the single point $(\infty, \infty)$ about which we as usual have 
$(x,y) \sim (\epsilon^{-2}, \epsilon^{-3})$ so the ramification index is also 4.
We subsequently have a square-tiling of the torus with passport
{\tiny $\begin{Bmatrix}
4\\
2,2\\
4\\
\end{Bmatrix}$}, which satisfies \eqref{RR}.
We emphasize this is not the conifold theory (cf.~\cite{He:2012js}), which has 1 single pair of black-white nodes each of valency 4 as well as 2 inequivalent squares.
What we have here is an unbalanced dimer model since we have 1 black node and 2 white nodes.
Nevertheless, the ramification structure is that of the conifold if we switched the points 1 and $\infty$.
While this is doable on the target $\IP^1$ by a simple $SL(2;\IZ)$ transformation, this permutation would result in a different form of the elliptic curve and hence would not be useful for ABC-triples, as dictated in Theorem \ref{thm:KS}.
For reference, multiplying the curve through by 4 and scaling $y$, we have that $(g_2,g_3) = (-4D,0)$, so the $j$-invariant here is simply 1728 using \eqref{j}.

\section{Points on the Curve from ABC-triple}\label{ap:curve}\setall
In this appendix we show that for every $r\in \overline{\mathbb{Q}}\setminus\{0,1,\infty\}$ there is precisely one positive cube-free integer $D$ such that the two Belyi maps given by
\begin{align}
\beta_{\pm}\colon&E_{D}\to\mathbb{P}^1\left(\mathbb{C}\right) \ ; \qquad
\left(x,y\right)\mapsto \frac{y \pm D}{\pm2D} = \frac{\pm y +D}{2D}
\end{align}
on the Type 2 elliptic curve $E_D$ defined by
\begin{equation}
\label{definingEquation_xy}
y^2 = x^3 + D^2
\end{equation}
admit rational preimages $\left(x,y\right)$ of $r$. We give explicit formulae for $D$, $x$ and $y$ and show that $\left(x,y\right)$ is uniquely determined up to the sign of $y$.
For the sake of generality, we will drop the condition that $a,b,c>0$ and consider instead any triple of coprime integers $a$, $b$, $c$ satisfying $a+b+c=0$.

\subsection{Computing the Preimages}
First of all, let $\left(x,y\right)$ be a rational point on $E_D$. Write
$x=\frac{s}{m}$ and  $y=\frac{t}{n}$ with $\gcd(s,m)=1=\gcd(t,n)$ and $m,n>0$.
Plugging this into \ref{definingEquation_xy} yields
\begin{gather}
\label{definingEquation_st}
t^2 m^3 = n^2s ^3 + m^3 n^2 D^2
\implies
\begin{cases}
	n^2 \left(s^3+m^3 D^2\right) = t^2 m^3 &\implies n^2\vert m^3\\
	m^3 \left(t^2-n^2 D^2\right) = n^2 s^3 &\implies m^3\vert n^2
\end{cases}
\end{gather}
Therefore $m^3=n^2=d^6$ for some $d\in\mathbb{Z}\setminus\{0\}$ and we can always write
\begin{align}
x=\frac{s}{d^2} \ , \qquad y=\frac{t}{d^3} \ .
\end{align}

Now let $a, b, c \in \mathbb{Z}\setminus\{0\}$ coprime such that 
\begin{equation}
a+b+c=0
\end{equation}
with prime decompositions
\begin{align}
a=\pm\prod_{p}p^{\alpha_p} \ , \quad
b=\pm\prod_{p}p^{\beta_p} \ , \quad
c=\pm\prod_{p}p^{\gamma_p} \ .
\end{align}
Further suppose that (note the difference in sign stems from the fact that the sign of $c$ has switched compared to the main text)
\begin{equation}
\beta\left(x,y\right)\equiv\frac{y+D}{2D} = -\frac{c}{b}
\end{equation}
or, using the notation introduced above,
\begin{align}
 2Dc =-\left(y+D\right)b \quad
\Leftrightarrow \qquad \frac{t}{d^3 D} = \frac{c-a}{c+a} \ .
\end{align}
Note that
\begin{gather}
\gcd(c+a,c-a)=\gcd(2c,b) =
\begin{cases}
2 $, if b is even$ \\
1 $, if b is odd$ \ .
\end{cases}
\end{gather}
We treat both cases separately.

\paragraph{Case of $b$ Even: }
Clearly we can write
\begin{align}
\label{td3D(n)even}
t = \frac{c-a}{2}n \ , \qquad
d^3 D = \frac{c+a}{2}n
\end{align}
for some $n\in\mathbb{Z}\setminus\{0\}$. Now we exploit \ref{definingEquation_xy} to find
\begin{equation}
\label{s(n)even}
s^3 = -acn^2 \ .
\end{equation}
The requirement that the RHS be a perfect cube serves as a constraint on $n$. Define 
\begin{equation}
m := \prod_p p^{\overline{\alpha_p+\gamma_p}} \ ,
\end{equation}
where the overline $\bar\cdot$ denotes reduction mod 3. By coprimality of $a$ and $c$, $\alpha_p$ and $\gamma_p$ cannot both be nonzero for any given prime $p$. We therefore have 
\begin{align}
m = \prod_p p^{\overline{\alpha_p}+\overline{\gamma_p}}
  = \prod_p p^{\overline{\alpha_p}} \cdot \prod_p p^{\overline{\gamma_p}}
  \equiv \widetilde{a}\cdot \widetilde{c} \ ,
\end{align}
where the tilde $\widetilde{\cdot}$ denotes the reduction of prime factor multiplicities mod 3. Note that $m$ is the smallest positive integer such that $acm^2$ is a perfect cube. We can now find a nonzero integer $k$ such that
\begin{equation}
n=mk^3 \ .
\end{equation}
Next, we plug this expression into \ref{td3D(n)even}:
\begin{align}
\label{td3D(mk)even}
t = \frac{c-a}{2}mk^3\ , \qquad
d^3 D =  -\frac{b}{2}mk^3 \ .
\end{align}
Since we require $D$ to be a cube-free integer, $d^3$ must cancel all cubes on the RHS. Define
\begin{equation}
b^\prime := \prod_{p>2}p^{\beta_p} \ , \quad 
v :=
\begin{cases}
2^2 \widetilde{b^\prime} & \mbox{ if }  \overline{\beta_2} = 0\\
\widetilde{b^\prime} & \mbox{ if } \overline{\beta_2} = 1\\
2 \widetilde{b^\prime} & \mbox{ if } \overline{\beta_2} = 2 \ .
\end{cases}
\end{equation}
Just as before, $v$ is the smallest positive integer such that $b/2v$ is a perfect cube. We now can express $d$ and $D$ in terms of known quantities (note that $m$ only contains prime divisors of $a$ and $c$ and therefore, by coprimality, $\gcd(m,b)=1$):
\begin{align}
d = \mp\sqrt[3]{\frac{b}{2v}}k \, \qquad
D = \pm vm \ .
\label{D_even}
\end{align}
As a final ingredient, recall \ref{s(n)even} and rewrite:
\begin{equation}
s = -\sqrt[3]{acm^2}k^2 \ .
\end{equation}
We are finally in a position to state explicit formulae for $x$, $y$:
\begin{align}
x = \frac{s}{d^2} = \sqrt[3]{\frac{-4acm^2 v^2}{b^2}} =  \sqrt[3]{\frac{-4acD^2}{b^2}}\ , \quad
y = \frac{t}{d^3} = \mp\frac{\left(c-a\right)mv}{b} = \frac{\left(a-c\right)D}{b}
\end{align}
where we identified $D$ in the last step.

\paragraph{Case of $b$ Odd: }
In close analogy to the procedure in the case of even $b$ write
\begin{align}
\label{td3D(n)odd}
t = \left(c-a\right)n \ , \qquad
d^3 D = \left(c+a\right)n
\end{align}
for $n$ a nonzero integer. \ref{definingEquation_xy} then implies
\begin{equation}
\label{s(n)odd}
s^3 = -4acn^2 \ .
\end{equation}
Decompose $4ac$ as
\begin{equation}
4ac = 2^{2+\alpha_2+\gamma_2}\cdot \prod_{p>2} p^{\alpha_p+\gamma_p} = 2^{2+\alpha_2 +\gamma_2}\cdot a^\prime c^\prime \ .
\end{equation}
The smallest positive integer $\ell$ such that $4ac\ell^2$ is a perfect cube is given by
\begin{equation}
\ell = 
\begin{cases}
2^2 \widetilde{a^\prime}\widetilde{c^\prime} &$, if $ \overline{\alpha_2 + \gamma_2}=0\\
\widetilde{a^\prime}\widetilde{c^\prime} &$, if $ \overline{\alpha_2 + \gamma_2}=1\\
2\widetilde{a^\prime}\widetilde{c^\prime} &$, if $ \overline{\alpha_2 + \gamma_2}=2
\end{cases}
\end{equation}
and the general form of $n$ is
\begin{equation}
n=\ell k^3 \qquad k\in\mathbb{Z}\setminus\{0\} \ .
\end{equation}
Upon usage of \ref{td3D(n)odd} we obtain
\begin{align}
d^3 D =\left(c+a\right)\ell k^3 =-b\ell k^3 \ .
\end{align}
Again, $d^3$ must cancel the cubes on the RHS:
\begin{align}
d = \mp \sqrt[3]{\frac{b}{\widetilde{b}}} k \ , \qquad
D = \pm \widetilde{b}l \ .
\label{D_odd}
\end{align}
Retracing our steps and repeating the calculations done above we arrive at
\begin{align}
x =\frac{s}{d^2}=\sqrt[3]{\frac{-4ac (\ell \widetilde{b})^2}{b^2}} = \sqrt[3]{\frac{-4acD^2}{b^2}} \ , \quad
y =\frac{t}{d^3}=\mp \frac{\left(c-a\right)\ell\widetilde{b}}{b} = \frac{\left(a-c\right)D}{b} \ ,
\end{align}
which has the exact same form as the corresponding results for even $b$.

\paragraph{Invariance of $D$: }
In fact, even the formula for $D$ does not depend on whether or not $b$ is even. Compare \ref{D_even} and \ref{D_odd}:
\begin{align}
D_{\text{even}}&=\pm
\begin{cases}
2^2\cdot\widetilde{a b^\prime c} &$ , if $ \overline{\beta_2} = 0\\
\widetilde{a b^\prime c} &$ , if $ \overline{\beta_2} = 1\\
2\cdot\widetilde{a b^\prime c} &$ , if $ \overline{\beta_2} = 2
\end{cases}\ , \quad
D_{\text{odd}}&=\pm
\begin{cases}
2^2\cdot\widetilde{a^\prime b c^\prime} &$ , if $ \overline{\alpha_2+\gamma_2} = 0\\
\widetilde{a^\prime b c^\prime} &$ , if $ \overline{\alpha_2+\gamma_2} = 1\\
2\cdot\widetilde{a^\prime b c^\prime} &$ , if $ \overline{\alpha_2+\gamma_2} = 2
\end{cases} \ .
\end{align}
Note that, by coprimality, exactly one of the numbers $a$, $b$, $c$ is even; therefore $a^\prime = a$, $c^\prime = c$ for $b$ even and $b^\prime=b$ for $b$ odd, and furthermore:
\begin{equation}
\mu_2\left(a,b,c\right):= \overline{\alpha_2+\beta_2+\gamma_2} = 
\begin{cases}
\overline{\beta_2} &$ , if $b$ even$\\
\overline{\alpha_2+\gamma_2} &$ , if $b$ odd$ \ .
\end{cases}
\end{equation}
Therefore, $D$ can always be computed as:
\begin{equation}
D=\pm
\left\{
\begin{array}{ll}
  2^2 \cdot \widetilde{a^\prime b^\prime c^\prime} &$ , if $ \mu_2\left(a,b,c\right)=0
  \\
  \widetilde{a^\prime b^\prime c^\prime} &$ , if $ \mu_2\left(a,b,c\right)=1\\
  2 \cdot \widetilde{a^\prime b^\prime c^\prime} &$ , if $ \mu_2\left(a,b,c\right)=2\\
\end{array}
\right\} = \pm \widetilde{\left(\frac{abc}{2}\right)} \ .
\end{equation}


\subsection{Arithmetic on $E_D$}
Next, we assemble some results on the arithmetic properties of the curves arising from ABC-triples in the described fashion. Let $E_D\left(\mathbb{Q}\right)$ denote the Mordell-Weil (MW) group of rational points on $E_D$ with the usual group operation. By the Mordell-Weil theorem, this is a finitely generated abelian group and we can write
\begin{equation}
E_D\left(\mathbb{Q}\right) \cong \mathbb{Z}^r \times E_D\left(\mathbb{Q}\right)_\text{tors}
\end{equation}
where $r$ is a non-negative integer called the {\bf rank} and $E_D\left(\mathbb{Q}\right)_\text{tors}$ is a finite abelian group.
Let us make some remarks and computations regarding this MW group for our $E_D$.

It turns out that we can determine the torsion part $E_D\left(\mathbb{Q}\right)_\text{tors}$ by recall the following two general theorems \cite{LN,Maz} (cf.~a nice exposition in \cite{RS})

{\bf THEOREM}
  {\bf (Nagell-Lutz)}
Let $E$ be the elliptic curve defined by $y^2=x^3+\alpha x+\beta$ where $\alpha,\beta\in\mathbb{Z}$. Every affine (i.e., non-zero) torsion point $(x,y)$ satisfies:\\
(i)  $x,y\in\mathbb{Z}$ \mbox{ and } (ii)  $y=0 \lor y \vert 4\alpha^3+27\beta^2$ \ .

{\bf THEOREM}
  {\bf (Mazur)}
Let $E$ be any elliptic curve over $\mathbb{Q}$. Then $E(\mathbb{Q})_\text{tors}$ is isomorphic to one of the following 15 groups:
\begin{enumerate}
\item $\mathbb{Z}/{m\mathbb{Z}}$ where $m\in\{1,2,3,4,5,6,7,8,9,10,12\}$;
\item $\mathbb{Z}/2\mathbb{Z} \times \mathbb{Z}/2m\mathbb{Z}$ where $m \in \{1,2,3,4\}$ \ .
\end{enumerate}

Now specialize to the familiar case of $E_D$. For brevity, let $T_D :=  E_D(\mathbb{Q})_\text{tors}$. Note that in the vernacular of the Nagell-Lutz theorem, $\alpha=0$ and $\beta=D^2$. The key observation is that each of these curves has two affine integer inflection points, namely $P_\pm=(0,\pm D)$. As is easily checked, $2\cdot P_\pm = -P_\pm = P_\mp$, which means that $P_\pm$ generates a subgroup $S\mathrel{\unlhd}T_D$ which is isomorphic to $\mathbb{Z}/3\mathbb{Z}$. By Lagrange's theorem, the only possible torsion groups which admit such subgroups are:
\begin{enumerate}
\item $T\cong\mathbb{Z}/3$, in which case $P$ generates $E(\mathbb{Q})_\text{tors}$
\item $T\cong\mathbb{Z}/3n\mathbb{Z}$ where  $n\in\{2,3,4\}$ and $S\cong\langle n\rangle$; $\left[T\colon S\right]=n$
\item $T\cong\mathbb{Z}/2\mathbb{Z}\times\mathbb{Z}/6\mathbb{Z}$ where $S\cong\langle (0,2) \rangle$; $\left[T\colon S\right]=4$
\end{enumerate}
In the latter two cases $P$ is a division point, i.e. there exists $Q\in E(\mathbb{Q})$ such that $k\cdot Q=P$ for some integer $k>1$. More precisely, $P$ is 3-divisble if $T_D\cong \mathbb{Z}/9\mathbb{Z}$ and at least\footnote{As is easily seen, if $T\cong\mathbb{Z}/12\mathbb{Z}$ then $P$ is 4-divisible, hence 2-divisble.} 2-divisible in the remaining cases. In particular, $Q=(a,b)$ is itself a torsion point and therefore, by the Nagell-Lutz theorem, $a,b\in\mathbb{Z}$. 

\paragraph{$P$ is 2-divisible: }
Suppose there were an integral point $Q=(a,b)$ such that $2\cdot Q=P$. Certainly $b\neq0$ since otherwise $2\cdot Q=0$ as is easily seen\footnote{Alternatively, note that $b=0$ would imply $a^3 =-D^2$; in particular, $D$ would have to be $\pm 1$ or a third power, as discussed in the ensuing text.}. Straightforward calculation shows that the general form of the tangent to $E_D$ at $Q$ is 
\begin{equation}
\mu(t) = 
\begin{pmatrix}
a+2bt\\
b+3a^2t
\end{pmatrix}
\end{equation}
This line intersects $E_D$ again at $t_0 = \frac{3a(3a^3-4b^2)}{8b^3}$
and switching the sign of the second coordinate in $\mu(t_0)$ yields the result:
\begin{equation}
2\cdot Q = 
\begin{pmatrix}
\label{2Q}
a+\frac{3a}{4b^2}(3a^3-4b^2) \\
-b-\frac{9a^3}{8b^3}(3a^3-4b^2)
\end{pmatrix}
\end{equation}
In order for $2\cdot Q$ to equal $P$, the first coordinate must vanish:
\begin{align}
0 &= 4ab^2 +9a^4 - 12ab^2=a(9a^3-8b^2)=a(a^3-8D^2)
\end{align}
where we used $b^2 = a^3 + D^2$ in the last step. This implies that $D^2$ is a third power, which is absurd unless $D=\pm1$ (note that $a\neq0$ since otherwise $Q=\pm P$).

\paragraph{The Case $D=\pm1$: }
As is clear from the above, in that case $a=2$ and by definition of $E_D$, $b\in\{-3,3\}$. Requiring the second coordinate in \ref{2Q} to be $D$ necessitates
\begin{align}
0&=b^4 +D b^3-36b^2+216=(b\mp3)(b\pm6)(b^2\mp2b+12)
\end{align}
the rational roots of which are $\mp6$ and $\pm3$, which shows that $b$ and $D$ have the same sign. We arrive at
$Q_\pm=
\begin{pmatrix}
2\\
\pm3
\end{pmatrix}
$. A straightforward calculation using the fact that $2\cdot Q_\pm = P_\pm$ shows that
\begin{equation}
3\cdot Q_\pm = P_\pm+Q_\pm = 
\begin{pmatrix}
0\\
\pm 1
\end{pmatrix}+
\begin{pmatrix}
2\\
\pm 3
\end{pmatrix}=
\begin{pmatrix}
-1\\
0
\end{pmatrix} =: R
\end{equation}
Suppose there were yet another torsion point $S= (s,t) \in E_{\pm1}$. By the second part of Nagell-Lutz, $t$ divides 27 and we have four possibilities:
\begin{equation}
{s}^3 = 
\begin{cases}
1-1=0 &\text{if }t=\pm1 \\
9-1=2^3&\text{if }t=\pm3\\
81-1=2^4\cdot5&\text{if }t=\pm9\\
729-1= 2^3\cdot7\cdot13&\text{if }t=\pm27
\end{cases}
\end{equation}
Clearly, the only torsion points are the ones we have already found, namely $Q_\pm$, $P_\pm$, $R$ and $\mathcal{O}$. Since
\begin{equation}
2\cdot P_\pm = -P_\pm = P_\mp
\end{equation}
we have
\begin{align}
  2\cdot Q_\pm = P_\pm \ , \quad
  3\cdot Q_\pm = R \ , \quad
  4\cdot Q_\pm = P_\mp \ , \quad
  5\cdot Q_\pm = Q_\mp \ .
\end{align}
Therefore $T_{\pm1}\cong\mathbb{Z}/6\mathbb{Z}$, generated by $Q_\pm$.

\paragraph{$P$ is 3-divisible: }
In analogy to the above, suppose there were a (necessarily integral) point $Q=(a,b)\neq P$ such that $3\cdot Q=P$. Recall \ref{2Q}, which we may use because $b\neq0$ (otherwise $2\cdot Q=0$), and for notational simplicity denote $2\cdot{Q} =: (A,B)$. Note that since $a\neq0$ and 
\begin{equation}
3a^3-4b^2=-(b^2+3D^2)<0
\end{equation}
we have $A\neq a$. This allows us to compute the first coordinate of $3\cdot Q$ as follows:
\begin{equation}
x=\left(\frac{B-b}{A-a}\right)^2-A-a
\end{equation}
Setting this to zero, cancelling denominators and multiplying with $(2b)^6$ one arrives at
\begin{align}
  \nn
0\overset{!}{=}&(2b)^8+54a^6(2b)^4-18a^3(2b)^6+729a^{12}-486a^9(2b)^2+81a^6(2b)^4 \\
&
-\left(162a^9(2b)^2-108a^6(2b)^4+18a^3(2b)^6+729a^{12}-729a^9(2b)^2+243a^6(2b)^4-27a^3(2b)^6\right)
\end{align}
which is easily checked to be equivalent to
\begin{align}
0&\overset{!}{=}(2b)^6-3^2a^3(2b)^4 +3^4a^9
=a^9-96D^2a^6+48D^4a^3+64D^6
=q^3-24q^2+3q+1
\end{align}
where we used $b^2=a^3+D^2$ in the first and introduced $q :=  a^3/4D^2$ in the second step. This last expression, however, has no rational roots, which immediately implies that there is no rational point $Q\neq \pm P$ such that $3\cdot Q$ lies on the $y$-axis. In particular, $P$ is not 3-divisible.

We therefore conclude that the torsion subgroup of $E_D(\mathbb{Q})$ is given by:
\begin{equation}
T\cong
\begin{cases}
\mathbb{Z}/6\mathbb{Z} &\text{   for } D^2=1\\
\mathbb{Z}/3\mathbb{Z} &\text{   otherwise \ .} 
\end{cases}
\end{equation}

\paragraph{Arithmetic of $abc$-Points: }
Next, we examine the relationship between points corresponding to ABC-triples in the described manner. First of all, note that relabeling $a$, $b$ and $c$ in any order does not change the triple we are considering. This means that \textit{a priori} we have six different ways of associating a point $(x,y)$ to a given triple (strictly speaking we have 12 different possibilities, since swapping the signs of $a$, $b$, $c$ does not change the triple either. However, this only amounts to changing the sign of $y$ and therefore does not yield any new points).
Since $D$ depends only on the product $abc$, it is manifestly independent of the labeling. However, $x$ and $y$ may change e.g., swapping $a$ and $c$ will change the sign of $y$. Denoting by $P_{\sigma}$ the point obtained after applying a permutation $\sigma\in S_3$ to $(a,b,c)$, we obtain the following points:
\begin{align}
\nn
&
P_{()}=
\begin{pmatrix}
x_1\\y_1
\end{pmatrix} =: P \ , \quad
P_{(ac)}=
\begin{pmatrix}
x_1\\-y_1
\end{pmatrix}=-P \ , \quad
P_{(abc)}=
\begin{pmatrix}
x_2\\y_2
\end{pmatrix} =: Q \\
&
P_{(ab)}=
\begin{pmatrix}
x_2\\-y_2
\end{pmatrix}=-Q \ , \quad
P_{(cba)} =
\begin{pmatrix}
x_3\\y_3
\end{pmatrix} =: R \ , \quad
P_{(bc)} =
\begin{pmatrix}
x_3\\-y_3
\end{pmatrix}=-R
\end{align}
where
\begin{align}
  \nn
  x_1=\sqrt[3]{\frac{-4acD^2}{b^2}} \ , \quad 
  x_2=\sqrt[3]{\frac{-4baD^2}{c^2}} \ ,  \quad
  x_3=\sqrt[3]{\frac{-4cbD^2}{a^2}} \ ,  \\
  y_1=\frac{\left(a-c\right)D}{b} \ , \quad
  y_2=\frac{\left(b-a\right)D}{c} \ , \quad
  y_3=\frac{\left(c-b\right)D}{a} \ .
\end{align}
Note that the $x_i$ are pairwise different. This means that, adopting the notation
\begin{equation}
\begin{pmatrix}
\sigma\\\tau
\end{pmatrix}=
\begin{pmatrix}
s\\t
\end{pmatrix}+
\begin{pmatrix}
S\\T
\end{pmatrix}
\end{equation}
we can once more employ the formula
\begin{equation}
\sigma = \left(\frac{T-t}{S-s}\right)^2 -s-S
\end{equation}
to check that $P-Q$, $Q-R$ and $R-P$ all lie on the $y$-axis. In this case, $\tau$ is given by
\begin{equation}
\tau = -t+\frac{T-t}{S-s}\cdot s
\end{equation}
and taking advantage of the useful identities
\begin{align}
\left(x_i-x_j\right)^3&= (2D)^2\cdot
\begin{cases}
\left(\frac{c-b}{cb}\right)^2\left(c^2-b^2\right)\text{  for } \{i,j\}=\{1,2\}\\
\left(\frac{a-c}{ac}\right)^2\left(a^2-c^2\right)\text{  for } \{i,j\}=\{2,3\} \ , \\
\left(\frac{a-b}{ab}\right)^2\left(a^2-b^2\right)\text{  for } \{i,j\}=\{1,3\}
\end{cases}
\end{align}
and
\begin{align}
y_i+y_j&= 2D\cdot
\begin{cases}
\frac{b^2-c^2}{bc}\text{  for } \{i,j\}=\{1,2\}\\
\frac{c^2-a^2}{ac}\text{  for } \{i,j\}=\{2,3\}\\
\frac{a^2-b^2}{ab}\text{  for } \{i,j\}=\{1,3\}
\end{cases}
\end{align}
we find that
\begin{equation}
P-Q=Q-R=R-P=
\begin{pmatrix}
0\\-D
\end{pmatrix}
\in E_D(\mathbb{Q})_\text{tors}
\end{equation}
Therefore $P$, $Q$ and $R$ agree up to torsion. Conversely, adding torsion elements to a given point does not change the triple it corresponds to. In particular, a point arising from an ABC-triple is never a torsion element.

Further questions beyond the scope of this work are whether or not these points actually \textit{generate} the torsion-free part of the Mordell-Weil group, how the quality of integer multiples of $abc$-points behave (Khadavi and Scharaschkin have some results on this) and, perhaps most interestingly, if there are any nontrivial arithmetic relations between points arising from different ABC-triples on the same curve.


\subsection{L-Function}
Whilst we are on the subject of arithmetic, it will be instructive to study the L-function associated to the elliptic curve $E_D$.
Let us consider the case of $E_D$ associated with the highest-quality $abc$-triple, viz., $D=3\times 23^2\times 109$.
The discriminant of $E_D$ is given by $\Delta=-2^4\times 3^7\times 23^8\times 109^4$, from which it follows that $2,3,23, 109$ are primes of bad-reduction.
Then we define
\begin{align*}
  a_p=\left\{
  \begin{array}{ll}
    p-N_p, & \hbox{$p= 2,3, 23,109$;}\\
    p+1-N_p, & \hbox{$p$\textrm{ other primes},} 
  \end{array}
  \right.
\end{align*}
where $N_p$ is the number of $\mathbb{F}_p$-points on the elliptic curve. Moreover we define $a_n$ for non-prime numbers $n$ as follows: $a_1=1$, $a_{p^k}=a_{p^{k-1}}a_p-pa_{p^{k-2}}$
for good primes $p$, and $a_{mn}=a_ma_n$ if $m$ and $n$ are relatively prime.
For our case, since $a_2=0$, $a_n$ vanishes if $n$ is even. The following list provides the  values of $a_n$ for the odd numbers less than 100:
\begin{align}
  \nn
  a_{2n+1} = \{ &
  1, 1, 1, 2, -2, 0, 3, 1, 1, 0, 2, 0, -4, -5, 1, 1, 0, 2, 1, 3, 1, 1,
  -2, 0, -3, 1, 0, 0, 0, 1, 2, \\
  & -4, 3, 1, 0, 1, 1, -4, 0, 21, 1, 1, 1, 1, 6, 1, 0, 1, 0
  \}
\end{align}
 Therefore the L-function  associated the given elliptic curve $E_D$ is explicitly expressed as
\begin{equation}
 \mathcal{L}(s,E_D)=\sum_{n=1}^\infty\frac{a_n}{n^s}=\frac{1}{1^s}+\frac{1}{3^s}+\frac{1}{5^s}+\frac{2}{7^s}-\frac{2}{9^s}+\cdots
\ , \quad s\in\mathbb{C} \ .
\end{equation}


\section{$E_D$: Lattice and Weierstrass Invariants}\label{ap:wp}\setall
In this appendix we examine $E_D$ from the complex analytic point of view. It is well known that any elliptic curve is conformally equivalent to a quotient $\mathbb{C}/\Lambda$ where
\begin{equation}\label{Lamgens}
\Lambda = \mathbb{Z}\langle 2\omega_1\rangle\oplus\mathbb{Z}\langle 2\omega_2\rangle
\end{equation}
denotes some lattice in the complex plane generated by two periods $2\omega_i$ which are linearly independent over $\mathbb{R}$ (we adhere to the convention, for the sake of the Weierstra\ss\ elliptic function, of using twice omega as the periods).

In the first step, we explicitly construct $\Lambda$ such that $E_D\cong \mathbb{C}/\Lambda$ \textit{over the rationals}. We then go on to identify all of these tori by allowing maps defined over $\overline{\mathbb{Q}}$.
Recall that, given a nonsingular elliptic curve $E$ defined by an equation of the form
\begin{equation}
y^2 = 4x^3+\alpha x +\beta
\end{equation}
there exists a lattice $\Lambda$ with Weierstra\ss\ constants
\begin{align}
g_2\left(\Lambda\right) = -\alpha \ , \quad g_3\left(\Lambda\right) = -\beta
\end{align}
if and only if the discriminant $\Delta := 4\alpha^3 + 27\beta^2$ is non-vanishing,
which in turn is equivalent to $E$ being nonsingular.

The isomorphism is then given by
\begin{align}
\label{WeierstrassIso}
\mathbb{C}/\Lambda &\to \mathbb{P}^{2}\left(\mathbb{C}\right) \ ; \qquad
\left[z\right] \mapsto
\begin{cases}
\left[\wp(z)\colon\wp^\prime(z)\colon1\right] &$, if $ \left[z\right]\neq0 \\
\left[0:1:0\right] &$, if $ \left[z\right]=0
\end{cases}
\end{align}
Taking as a point of departure the familiar defining equation for $E_D$,
\begin{equation}
y^2=x^3+D^2
\end{equation}
we first perform a simple substitution (defined over $\mathbb{Q}$):
\begin{align}
4v := y \ , \quad 4u := x
\end{align}
to find the equivalent Weierstra\ss\ form
\begin{equation}
v^2=4u^3+\left(\frac{D}{4}\right)^2 \ .
\end{equation}

In light of \ref{WeierstrassIso} and the well known differential equation for the Weierstra\ss\ $\wp$-function,
\begin{equation}
\label{WeierstrassDiffEq}
\left(\wp^\prime\right)^2 = 4\wp^3 - g_2 \wp -g_3 \ ,
\end{equation}
we are looking for a lattice $\Lambda$ which satisfies
\begin{align}
\label{WeierstrassVsD}
g_{2}\left(\Lambda\right) = 0 \ , \qquad
g_{3}\left(\Lambda\right) = -\left(\frac{D}{4}\right)^2 \ .
\end{align}
To this end let $\Lambda = \mathbb{Z}\langle 2\omega_1\rangle \oplus \mathbb{Z}\langle 2\omega_2\rangle$, as in \eqref{Lamgens},
where $\omega_1$ and $\omega_2$ denote the half-periods and we can write
\begin{equation}
\begin{pmatrix}
\omega_1 \\
\omega_2
\end{pmatrix}
=
\omega_1
\begin{pmatrix}
1 \\
\frac{\omega_2}{\omega_1}
\end{pmatrix}
=:
\omega_1
\begin{pmatrix}
1 \\
\tau
\end{pmatrix} \ .
\end{equation}
We can achieve $\tau\in \cH$, the upper half plane, by switching the generators if necessary.
Recall that $g_2$ is defined in terms of an Eisenstein series as follows:
\begin{align}
\nn
g_{2}\left(\Lambda\right) &= 60G_{4}\left(\Lambda\right) = 60\sum\left(2m\omega_1 + 2n\omega_2\right)^{-4} = 60\left(2\omega_1\right)^{-4}\sum\left(m + n\tau\right)^{-4} \\
&= 60\left(2\omega_1\right)^{-4}G_{4}\left(\Lambda_\tau\right) = \left(2\omega_1\right)^{-4}g_{2}\left(\Lambda_\tau\right) \ ,
\end{align}
where $(m,n)$ run over $\mathbb{Z}^2\setminus\{(0,0)\}$ and
\begin{equation}
\Lambda_\tau := \mathbb{Z}\langle1\rangle\oplus\mathbb{Z}\langle\tau\rangle \ .
\end{equation}
Therefore, $g_2\left(\Lambda\right)$ vanishes if and only if $g_2\left(\Lambda_\tau\right)$ does.
Using $j(\Lambda)=1728\frac{g_2^3}{g_2^3-27g_3^2}$ from \eqref{j},
we can restate this condition as
\begin{equation}
j(\Lambda_\tau)=0
\end{equation}
which is satisfied if and only if
\begin{equation}
\exists M=
\begin{pmatrix}
a&b\\c&d
\end{pmatrix}
\in \text{SL}_2(\mathbb{Z})\ \ \colon\ \tau = M\langle \zeta_6\rangle
\end{equation}
where $\zeta_6=\exp(\frac{2\pi i}{6})$ denotes the primitive sixth root of unity and $M\langle - \rangle$ is the usual action on $SL(2;\IZ)$ on complex numbers by fractional linear transformations.

Now, $\omega_1$ needs to be chosen appropriately in order to satisfy the second constraint in \ref{WeierstrassVsD}. In analogy to the above, we have
\begin{align}
\nn
g_{3}\left(\Lambda\right) &= 140G_{6}\left(\Lambda\right) = 140\left(2\omega_1\right)^{-6}G_{6}\left(\tau\right) = 140\left(2\omega_1\right)^{-6}G_{6}\left(M\langle\zeta_6\rangle\right) \\
&= 140\left(2\omega_1\right)^{-6} \left(c\zeta_6+d\right)^{6}G_{6}\left(\zeta_6\right)= \left(\frac{c\zeta_6+d}{2\omega_1}\right)^{6}g_3\left(\Lambda_0\right)
\end{align}
where $\Lambda_0$ is shorthand notation for the lattice generated by $\{1,\zeta_6\}$ and we used that $G_6$, when viewed as a function on $\cH$, is a modular form of weight 6.
Solving for $\omega_1$, plugging in \ref{WeierstrassVsD} and using
\begin{equation}
g_3(\Lambda_0) = \frac{\Gamma\left(\frac{1}{3}\right)^{18}}{\left(4\pi\right)^6} =: C^6 \text{ , } C\in\mathbb{R}_{>0}
\end{equation}
from \cite{AS}, gives
\begin{equation}
\omega_1 = \frac{C}{\sqrt[3]{4D}}(c\zeta_6+d)\zeta_{12}^{2k+1}
\end{equation}
and finally
\begin{equation}
\label{generalSolution}
\begin{pmatrix}
\omega_1 \\
\omega_2
\end{pmatrix}=\frac{C}{\sqrt[3]{4D}}\zeta_{12}^{2k+1}
\begin{pmatrix}
c\zeta_6+d\\
a\zeta_6+b
\end{pmatrix}
\end{equation}
where $k\in\{0,1,2,3,4,5\}$. Now we can specialize to more tractable cases, e.g.\ $k=5$ and $M=\II_2$:
\begin{equation}
\omega_1 =\frac{C}{\sqrt[3]{4D}} \overline{\zeta_{12}} \ ,
\qquad
\omega_2 = \frac{C}{\sqrt[3]{4D}}\zeta_{12} \ .
\end{equation}

In fact, if $k$ is left fixed, then varying $M$ leaves the lattice itself invariant, since two bases yield the same lattice if and only if they are related by a $\text{GL}_2(\mathbb{Z})$ action (here, ``$\text{GL}_2(\mathbb{Z})$ action'' means the usual action on a 2-vector of generators by matrix multiplication). Therefore we may choose $M=\II_2$ without loss of generality. Next, note
\begin{equation}
\zeta_{12}^{2k+1} = \zeta_{12}\cdot\zeta_6^k
\end{equation}
However, scalar multiplication of the generators with $\zeta_6^k$ also leaves the lattice invariant:
\begin{align}
\nn
\zeta_6\cdot
\begin{pmatrix}
1 \\
\zeta_6
\end{pmatrix}=
\begin{pmatrix}
\zeta_6 \\
\zeta_6^2
\end{pmatrix}&=
\begin{pmatrix}
\zeta_6 \\
\zeta_6-1
\end{pmatrix}=
\begin{pmatrix}
0 & 1 \\
-1 & 1
\end{pmatrix}\cdot
\begin{pmatrix}
1 \\
\zeta_6
\end{pmatrix}
\Rightarrow
\zeta_6^k \cdot
\begin{pmatrix}
1 \\
\zeta_6
\end{pmatrix}&=
\begin{pmatrix}
0 & 1 \\
-1 & 1
\end{pmatrix}^k\cdot
\begin{pmatrix}
1 \\
\zeta_6
\end{pmatrix}
\end{align}
Therefore, the lattice is, up to an overall scale factor, completely determined to be
\begin{equation}
\Lambda = \mathbb{Z}\langle \overline{\zeta_{12}}\rangle \oplus \mathbb{Z}\langle \zeta_{12}\rangle \ .
\end{equation}


\end{document}